\documentclass[twocolumn,prl,showpacs,floatfix]{revtex4-1}
\usepackage{graphicx}
\usepackage{epsfig}
\usepackage{rotating}
\usepackage{amsmath}
\usepackage{amsfonts}
\usepackage{amssymb}
\usepackage{enumerate}
\usepackage{longtable}
\usepackage{dcolumn}
\usepackage{bm}
\usepackage{physics}
\usepackage{verbatim}
\newcommand{\zncu}{ZnCu$_3$(OH)$_6$Cl$_2$ }

\begin{document}

\title{Structure Factors of the Kagome-Lattice Heisenberg antiferromagnets at finite temperatures}

\author{Nicholas E. Sherman$^1$ and Rajiv R. P. Singh$^1$}
\affiliation{$^1$Department of Physics, University of California Davis, CA 95616, USA }

\date{\rm\today}

\begin{abstract}
We compute the real-space spin correlations and
frquency and wave-vector resolved dynamic structure factors $S(\vec q,\omega)$ for the nearest-neighbor Kagome-Lattice
Heisenberg Antiferromagnet (KLHM) at finite temperatures using Numerical Linked Cluster Expansion (NLCE) method. A triangle-based
NLCE is used to calculate frequency moments of the dynamic structure factors in the thermodynamic limit, which show excellent
convergence for $T> J/4$. 
A Gaussian approximation and the fluctuation-dissipation relation is used to
to reconstruct the frequency dependence.
We find that some features of the low temperature KLHM structure-factors
begin to set in at temperatures of order $J$. 
Our results are in very good agreement with powder diffraction measurements reported earlier on the Herbertsmithite materials \zncu.
However, the calculated properties differ from the low temperature ($T\approx J/100$)
experimental measurements in one important regard. 
In line with the experimental observations, the spectral weight has a diffuse nature, which is 
predominantly spread along the extended Brillouin-Zone boundary. 
However, the maximum intensity is found in our calculations to be at the $K$ point 
of the extended Brillouin Zone in contrast
to the low temperature experiments, where it is at the $M$ point. We suggest that experiments should be done at various
temperatures to look for such a crossover of the maximum from the $K$ point to the $M$ point. In the absence of such a
crossover, the Herbertsmithite materials must differ from KLHM in a significant manner.
\end{abstract}


\maketitle
\section{Introduction}
The spin-half nearest-neighbor Kagome Lattice Heisenberg Model (KLHM) is one of the best studied models of quantum magnetism \cite{misguich,balents}. 
Recent computational studies have established a quantum spin-liquid ground state for the model, although
the full nature of the quantum spin-liquid phase and the existence of a spin-gap remains under debate \cite{dmrg,ran07,gapless,recent}.
While the breakthrough DMRG studies suggested a gapped spin-liquid with a robust spin-gap
of order or larger than a tenth of the exchange constant $J$ \cite{dmrg}, several recent studies suggest a Dirac spin-liquid with
gapless excitations \cite{recent}. 

On the experimental front, the Herbertsmithite materials ZnCu$_3$(OH)$_6$Cl$_2$ 
have been celebrated as possibly
a nearly ideal realization of a Kagome antiferromagnet \cite{imai-lee,klhm-e}. The lack of structural distortions and magnetic isolation of copper based
Kagome planes by intervening non-magnetic zinc planes, makes these materials well suited to the exploration
of the rich quantum spin-liquid physics in these systems. Recent neutron-scattering \cite{neutron12} and NMR \cite{Fuscience} measurements on
large single-crystal materials find no evidence for magnetic order down to temperatures many orders of
magnitude below the exchange energy scale. While the as obtained neutron-scattering spectra clearly shows
gapless excitations, a recent analysis of the spectra taking into account the anti-site copper impurities in the
zinc planes, assigns the low energy scattering entirely to these impurities \cite{han16}. The authors conclude a gap of order 
$J/20$ for the KLHM, in agreement with the eariler NMR study and 
somewhat below the DMRG calculations of Yan {\it et al} \cite{dmrg}.

In this work, we aim to calculate the real-space spin correlations and
strcture factors at intermediate temperatures, where the
gap issue is not relevant. Our main goal is to benchmark the structure factor for the KLHM, 
in a temperature regime where they can be accurately calculated in the thermodynamic limit, and thus be directly
compared to the experiments. Despite the many theoretical studies, this quantity has not been calculated before
apart from a high temperature expansion study at selected wavevectors \cite{elstner},
and is important for addressing the question of how good the Heisenberg Model is for these materials \cite{nmr-theory}.
The static structure factors, we calculate, should be very accurate down to the lowest temperatures
studied. The prominent features of the wave-vector dependence of the structure factor begin to develop at relatively high temperatures
of order $J$. The frequency dependence is obtained through the Gaussian approximation,
which should be a good approximation for the short-time dynamics \cite{gelfand}.

We find that, in the static structure factors as
well as the low-energy structure factors, the intensity is mostly spread near the extended Brillouin-Zone boundary once
the temperature is below the exchange energy sclae $J$. 
However, we find that
the intensity peaks at the K-point in the extended Brillouin Zone. This is in contrast to the low
temperature experimental observation at $T=J/100$, where the peak is at the M point \cite{neutron12,sachdev,hao}. We note that the finite
size calculation of Shimokawa and Kawamura \cite{kawamura}  also found a crossover of the maximum in the structure factor from
the K point to the M point at $T\approx J/100$. Thus, our results are fully consistent with their studies.
Our work suggests that measuring the temperature dependence of the structure factors as a function of
temperature can help clarify how good the KLHM is for these materials and determine an important crossover energy scale.

\section{Model and Methods}
We consider the Heisenberg model with Hamiltonian:
\begin{equation}
{\cal H}= J\sum_{\langle i,j \rangle} (S_i^x S_j^x + S_i^y S_j^y+S_i^z S_j^z),
\end{equation}
where the sum runs over all nearest-neighbor bonds of the Kagome lattice.
The $S_i^\alpha$ ($\alpha=x,y,z$) represent spin-half operators associated with
the spin at site $i$.

The Kagome Lattice consists of 3-sublattices, which we can label by $a,b=1,2,3$. A site
on the Kagome lattice has location:
\begin{equation}
{\vec r} = n_1 \vec R_1 + n_2 \vec R_2 +\vec l_a
\end{equation}
where $\vec R_1$ and $\vec R_2$ are the lattice translational vectors of the underlying triangular Bravais-lattice,
$n_1$ and $n_2$ are integers, and $\vec l_a$ for $a=1,2,3$ are the $3$ basis vectors in a unit cell.
One explicit representation (taking nearest neighbor distance of unity) is:
$$\vec R_1= 2 \hat x, \qquad \vec R_2= \hat x + \sqrt{3} \hat y,$$
with basis vectors,
$$\vec l_1=0, \qquad \vec l_2=\hat x, \qquad\vec l_3 ={1\over 2}\hat x + {\sqrt{3}\over 2} \hat y.$$

Neutron scattering measures the scattering cross-section resolved by momentum transfer $\vec q$ and energy transfer $\omega$
(we set $\hbar=1$).
Let us begin with the correlations in time $t$ instead of energy transfer $\omega$.
For momentum transfer $\vec q$, the dynamic structure factor is a $3\times 3$ matrix:
\begin{align}
S_{ab}(\vec q,t) = & \sum_{n_1,n_2} \left[\langle e^{-i H t} S_a(0,0)e^{i H t} S_b(n_1,n_2)\rangle\right. \\ \nonumber
\cross & \left.e^{-i \vec q \cdot (n_1 \vec R_1 + n_2 \vec R_2)}\right] e^{i \vec q \cdot (\vec l_a -\vec l_b)}
\end{align}
Fourier transforming in time gives
\begin{equation}
S_{ab}(\vec q,\omega ) = e^{i \vec q \cdot (\vec l_a -\vec l_b)} \sum_{n_1,n_2} S_{ab}(n_1,n_2,\omega)  e^{-i \vec q \cdot (n_1 \vec R_1 + n_2 \vec R_2)}
\end{equation}
where $ S_{ab}(n_1,n_2,\omega)$ is the time Fourier transform of $\langle e^{-i H t} S_a(0,0)e^{i H t} S_b(n1,n2)\rangle$, which by translational symmetry only
depends on vector distance given by $n_1$, $n_2$. The neutron scattering cross-section
is the sum over all $9$ matrix elements of the $S_{ab}$ matrix.

The equal-time correlation function is obtained by summing over all frequencies, and leads to the expression

\begin{align}
S_{ab}(\vec q) =& \sum_{n_1,n_2} \left[\langle S_a(0,0) S_b(n1,n2)\rangle\right. \\ \nonumber 
\cross& \left. e^{-i \vec q \cdot (n_1 \vec R_1 + n_2 \vec R_2)}\right] e^{i \vec q \cdot (\vec l_a -\vec l_b)} 
\end{align}

In this work, we calculate real-space spin-spin correlation functions $\langle S_a(0,0) S_b(n1,n2)\rangle $, as well as static and
dynamic structure factors $S_{ab}(\vec q)$ and $S_{ab}(\vec q,\omega )$ using the Numerical Linked Cluster method. 

\section{Numerical Linked cluster method}
The essence of the Numerical Linked Cluster (NLC) method is to express an extenive property $P$ 
for a large lattice ${\cal L}$ with $N$-sites as
\begin{equation}
P({\cal L})/N=\sum_c L(c)\times W(c).
\end{equation}
Note that given an intensive property $p$ such as spin-spin correlation function, one
can always construct an extensive property by defining $P=Np$.
Here, the sum over $c$ runs over all distinct linked clusters of the lattice ${\cal L}$.
$L(c)$ is called the lattice constant of the cluster c, and is the number of embeddings of the linked-cluster 
in the lattice per site. The quantity $W(c)$ is called the weight of the cluster and
is determined entirely by a calculation of the property on the finite cluster $c$ and all its sub-clusters. 
It is defined as
\begin{equation}
W(c)=P(c)-\sum_s W(s),
\end{equation}
where the sum over $s$ is over all proper subclusters of the cluster $c$. In a high temperature expansion, the
property $P(c)$ is expanded in powers of inverse temperature. In NLC, one carries out a calculation at a given
temperature by exact diagonalization of the finite system. 

For the Kagome lattice, it is useful to consider clusters made up of complete triangles only. It was found in Ref.~ \onlinecite{nlc-et}
that whereas an NLC based on bond or site based graphs starts to breakdown as soon as the high temperature expansion
diverges, the triangle based NLC converges down to much lower temperatures. We define the order of the calculation
by the clusters with largest number of triangles included in the calculation. We carry out complete calculations
for up to $7$ triangles or $7$th order.

\begin{figure}
\begin{center}
\includegraphics[width=\linewidth]{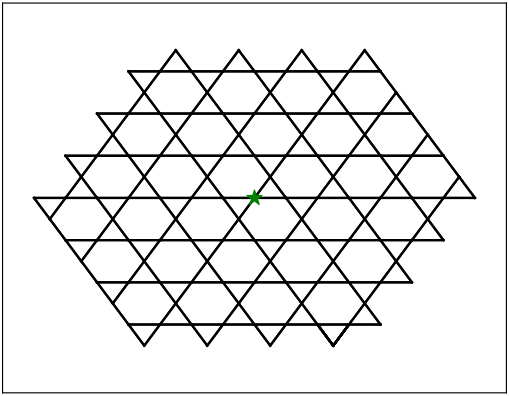}
\caption{\label{kagome} A section of the Kagome lattice. Shown are the sites whose
correlations with respect to a given site denoted by a star are calculated in the NLC expansion up to 7 triangles. }
\end{center}
\end{figure}

We first calculate the spin correlations in real space and then Fourier transform to get the correlations in
momentum space. The order of the NLC calculation limits the largest vector distance for which the correlations can
be non-zero. The range of real-space correlations studied in $7$th order NLC is depicted in Fig.~1. The figure
depicts all sites that are within $7$ triangles of a given site.
We will find that these distances are large enough that, at least at temperatures of interest
in this work, the correlations become very small well before the largest distances accessed in the study.

In order to calculate the real-space correlations $\langle S_a(0,0) S_b(n1,n2)\rangle$ for all $a$, $b$ and all relevant vector
distance $(n1,n2)$, we group them into symmetry distinct sets. The correlations will be identical for two vector
distances that are related by a symmetry of the Kagome-lattice. Up to $7$th order, there are
29 distinct vector distances. 
The following steps are needed to carry out the calculations:

\begin{itemize}

\item{} First we prepare a list of topological graphs, their subgraphs and
their lattice constants  up to some order $n$. An $n$-th order graph, with $n$ triangles will have
$N_s$ sites and $N_b$ bonds. The topology of the graph is fully specified by the connectivity or adjacency matrix of
the graph and this information is sufficient for calculating distance independent properties.

\item{} Coordinate-dependent embeddings of the graphs in the lattice are needed for calculating distance dependent
spin-spin correlation functions. For each topological graph, all possibe lattice embeddings are determined up to
symmetries of the lattice, together with their symmetry related count.

\item{} A list is prepared of all relevant 
vector distances $(a,b,n1,n2)$ divided into 29 distinct sets (for order $n=7$). 
That is, every vector between
pair of spins in all graphs must be in one and only one of the vector distances in the set.

\item{} For all embeddings, an identification for every pair of sites with one of the 29 elements is
made. That is, the vector-distance the pair
belongs to in the embedding, is determined. 

\item{} Using an exact diagonalization program, spin-spin correlations are determined for every pair of
spins of a topological graph.

\item{} Using the assignment of vector-distances to each pair,
the spin-spin correlation sum (and frequency moments) for all the $29$ distinct vector distances are calculated
for each graph. These define
the extensive properties for which weights can now be obtained by subgraph subtraction.

\item{} Once weights have been determined, summing over all topological graphs gives us the spin-spin correlation functions
for the infinite lattice.
This process gives us a sum over all symmetry related spin-spin correlations, per lattice-site. Knowing the
number of equivalent vectors in each case, the spin-spin correlation
between pairs of spins follows.

\item{}
Fourier Transforming the results gives us the wavevector dependence.
\end{itemize}

The frequency moments of $S_{ab}(n_1,n_2,\omega)$, can be written as thermal expectation values of commutation relations of on-site spin operators and the Hamiltonian. This ensures that linked-cluster expansion exists. In our NLC calculations,
we do not use the commutation relations. 
We have the exact eigenstates of the graph. Then, for spins at site $i$ and $j$, 
the frequency moments can be calculated from the expression,
\begin{equation}
    \rho_{ij}^k = \frac{1}{\mathcal{Z}}\sum_{n,m}e^{-\beta E_n} \langle n | S_i^z | m \rangle\langle m | S_j^z | n \rangle (E_m - E_n)^k
\end{equation}
Where $\{(|n\rangle,E_n)\}$ are the eigenvectors and eigenvalues of the Hamiltonian, $\beta$ is inverse temperature, and $\mathcal{Z}$ is the partition function. Due to the spin rotational symmetry of the Heisenberg model, it suffices to calculate the zz correlation functions.
We note that the zeroth moment is the equal time correlation function.  All results are Fourier transformed to wave-vector space
$\vec q$. To obtain the frequency dependence for any $\vec q$, we use the Gaussian approximation. 
For this, we first introduce the spectral density defined by
\begin{equation}\label{sd}
    \Phi(\omega) = \frac{1}{2}\left(1+e^{\beta\omega}\right)S(\omega)
\end{equation}
which is an even function in $\omega$, and also shares all of its even moments with $S$. Since this function is even, we assume that 
it is a gaussian with zero mean. Thus it is determined from its zeroth and second moment. 
After NLC and fourier transformation is performed for the zeroth and second moments, we construct $\Phi$ in the gaussian 
approximation and use equation (\ref{sd}) to determine $S_{a,b}(\vec{q},\omega)$.  The benefit of going through the spectral density is that this function is even in $\omega$, and so a gaussian with mean zero preserves the fluctuation-dissipation relations.

\section{Results}

We begin with the correlations in real space. There are 29 relevant vector distances, and Table-1 gives 
representative vectors for the first four of them. Note in particular that the vector distances with labels 0 and 1 correspond to on-site and nearest neighbor vector distances respectively.

\begin{center}
 \begin{tabular}{| c || c | c |} 
 \hline
 Label & $x\left(\frac{a}{2}\right)$ & $y\left(\frac{a\sqrt{3}}{2}\right)$ \\
 \hline
 0 & 0 & 0 \\
 1 & 1 & 1 \\
 2 & 0 & 2 \\
 3 & 4 & 0 \\
 \hline
\end{tabular}
\end{center}

\begin{figure}
\begin{center}
\includegraphics[width=0.49\linewidth]{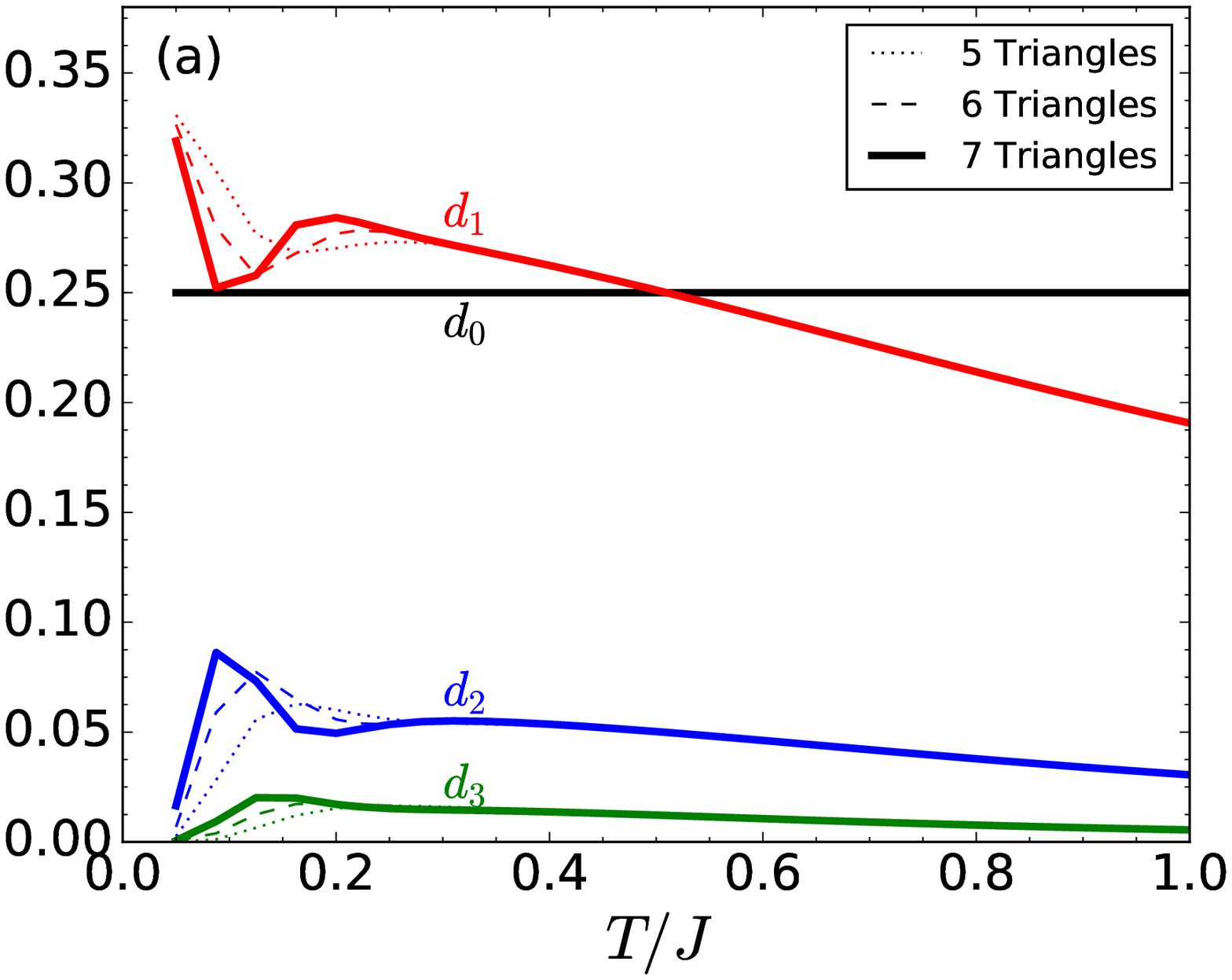}
\includegraphics[width=0.49\linewidth]{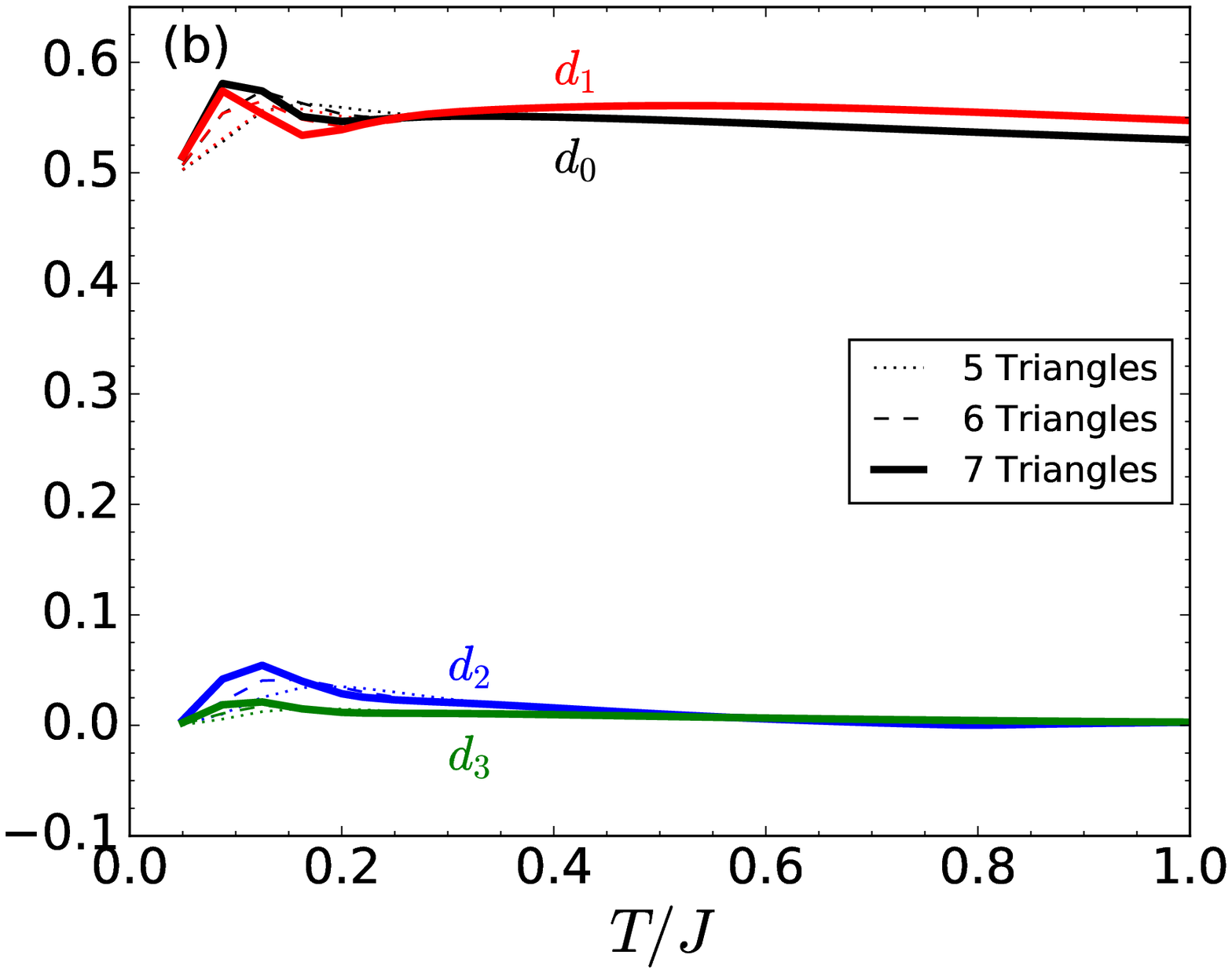}
\caption{\label{moments} The zeroth and second frequency moments of $S(\vec{r})$ for the first four vector distances in the Kagome AFM Heisenberg model. Plot (a) corresponds to the zeroth, and plot (b) corresponds to the second moment. }
\end{center}
\end{figure}

The zeroth and second moment for the first four vector distances are shown in Figure \ref{moments}. Here we show the result 
for 5th, 6th and 7th order. It is clear that convergence is excellent, with hardly much difference between 6th and 7th
orders down to a temperature of  $T=0.25J$.  Thus, for the remainder of this paper we only show results from calculations 
for our highest order that is 7th order (that is up to 7 triangle graphs), and do not show temperatures below $0.25J$ where
differences between $6$th and $7$th order begin to arise.

\begin{figure}
\begin{center}
\includegraphics[width=0.49\linewidth]{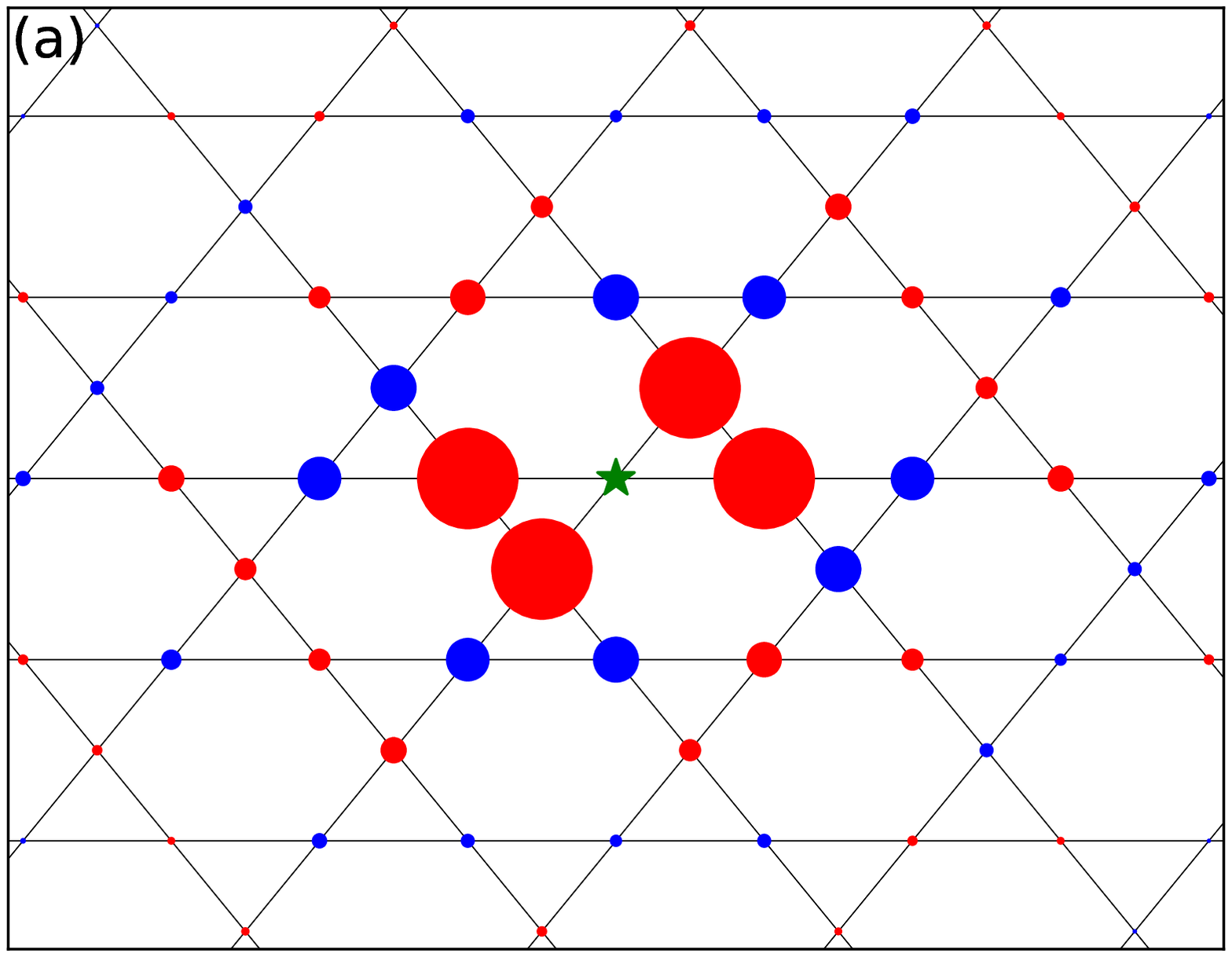}
\includegraphics[width=0.49\linewidth]{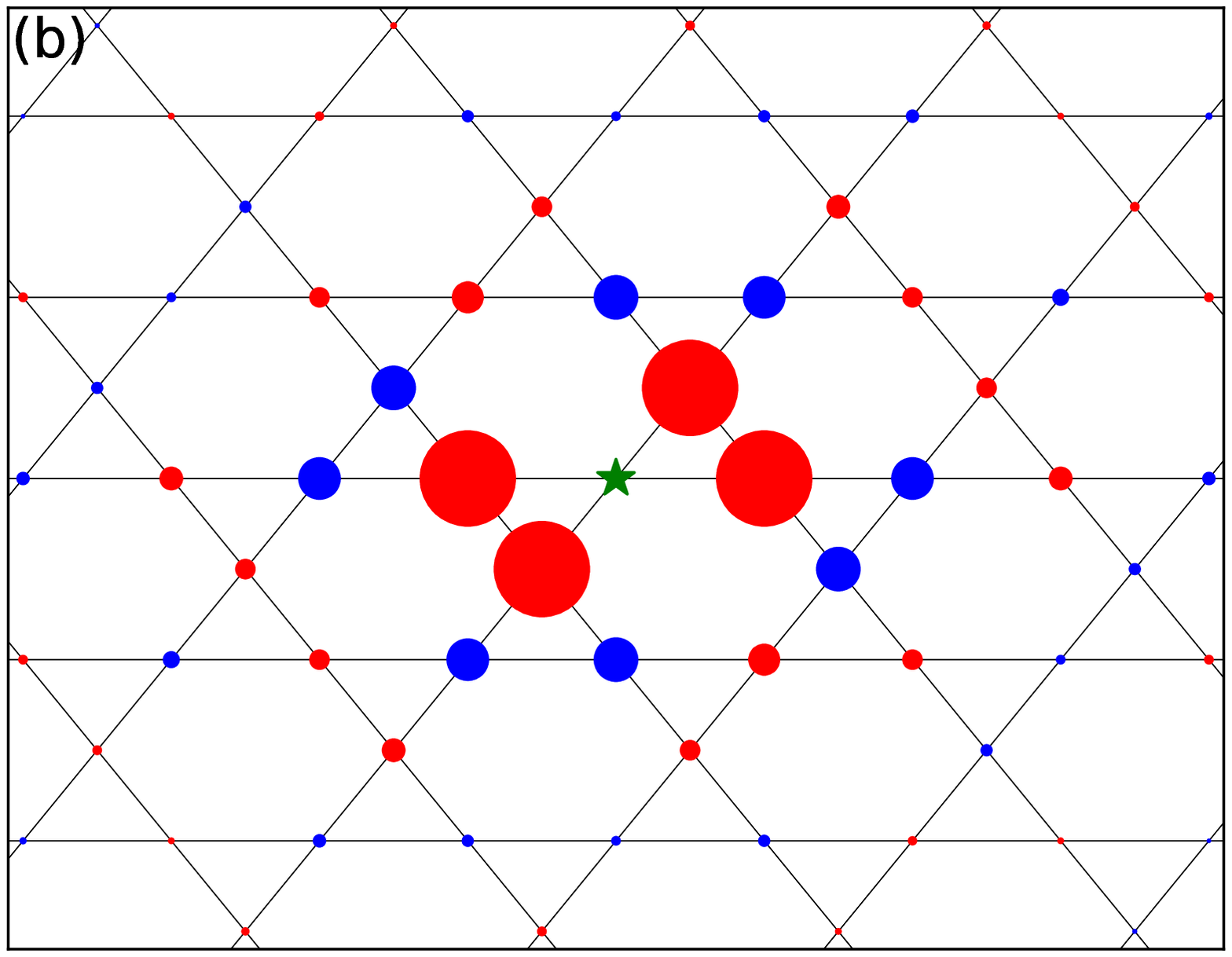}\\
\includegraphics[width=0.49\linewidth]{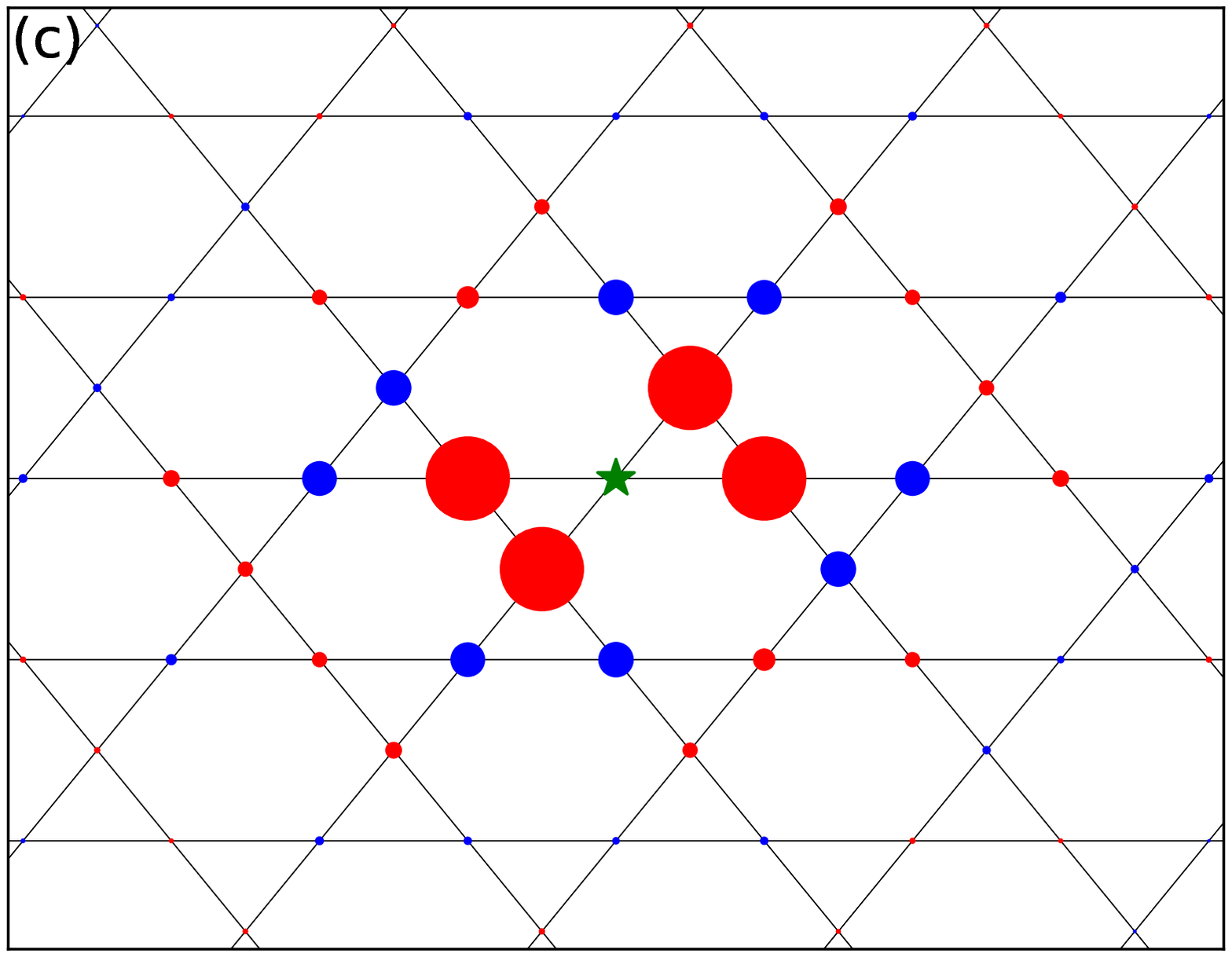}
\includegraphics[width=0.49\linewidth]{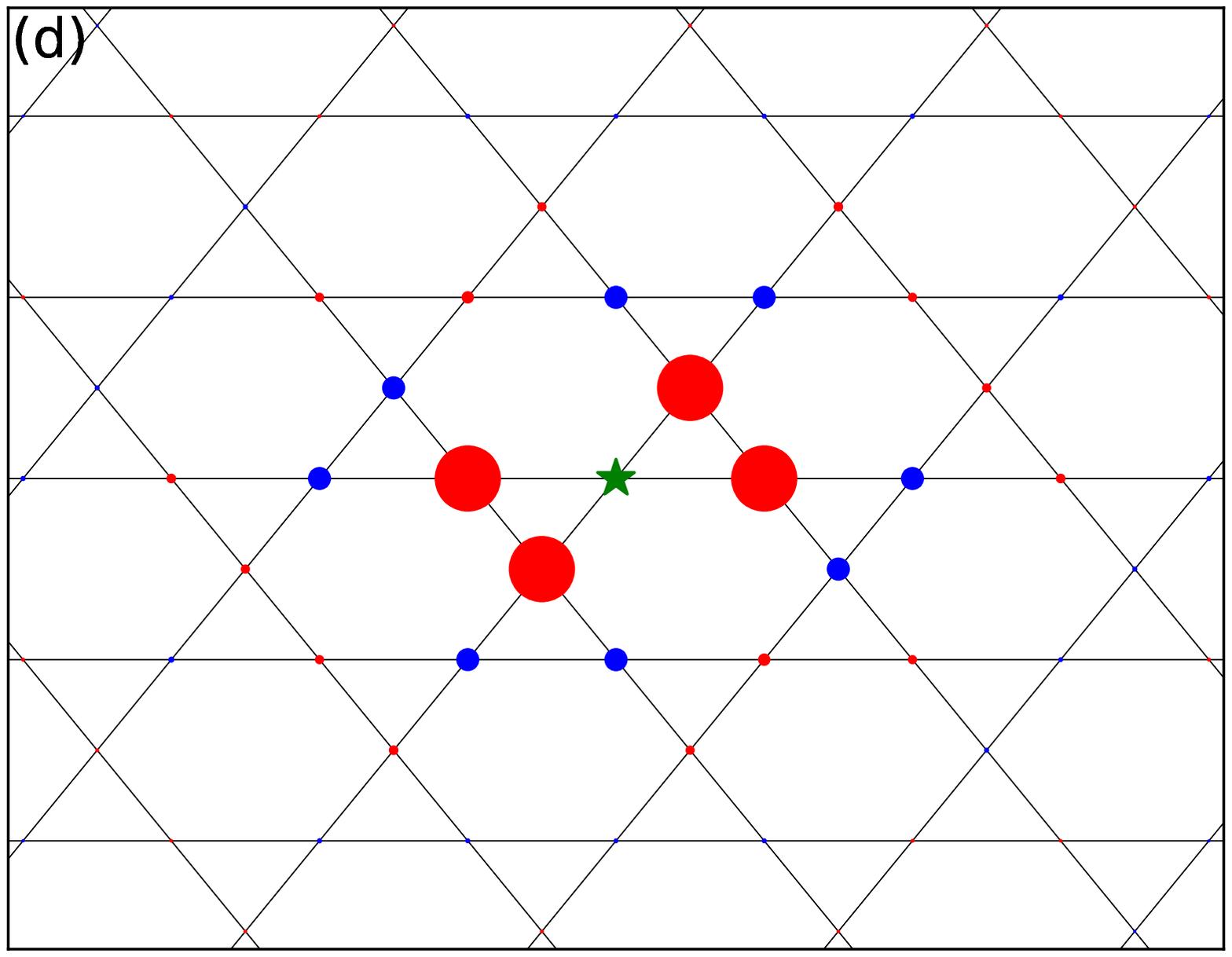}
\caption{\label{correlations} The equal time spin-spin correlations between sites in the Kagome AFM Heisenberg model.  The correlations are between each site, and the site depicted by a green star.  The area of the circle corresponds to the magnitude of the correlation, and the color red means a negative correlation, and blue is a positive correlation. The four plots depict the temperatures 0.25J, 0.5J, J, and 2J for plots (a), (b), (c), and (d) respectively.}
\end{center}
\end{figure}

Next we show pictorially the real-space correlations between spins on the Kagome lattice in Figure \ref{correlations}. In these plots we illustrate the equal-time, spin-spin correlation between a given site with the site labelled by a green star. 
On each site we draw a colored circle, whose area quantitatively illustrates the magnitude of the correlation. The color of the circle specifies the sign of the correlation, with red signifying a negative correlation and blue positive. We see that even down to a temperature of $0.25J$, significant correlations do not extend beyond a few lattice constants.

\begin{figure}
\begin{center}
\includegraphics[width=0.49\linewidth]{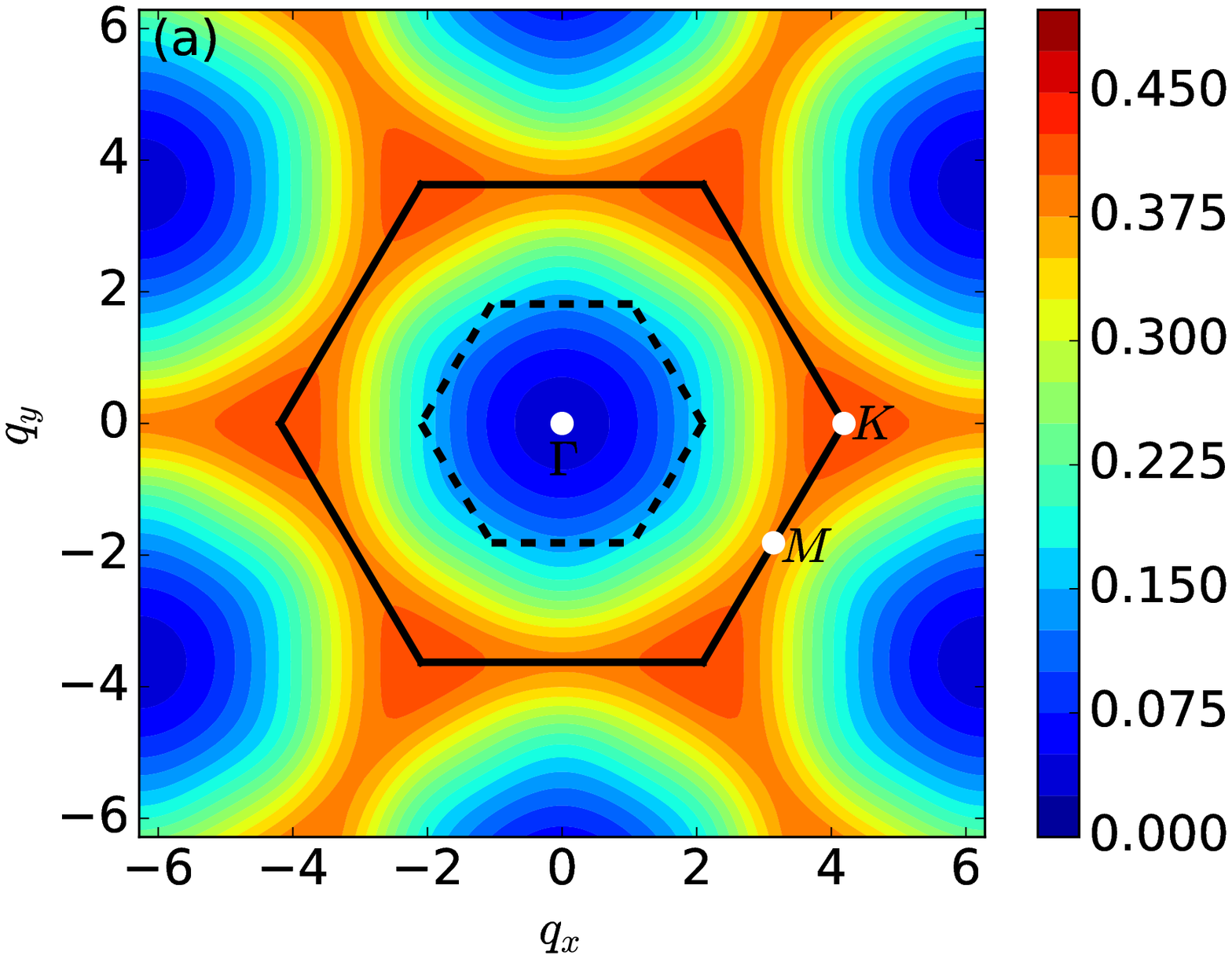}
\includegraphics[width=0.49\linewidth]{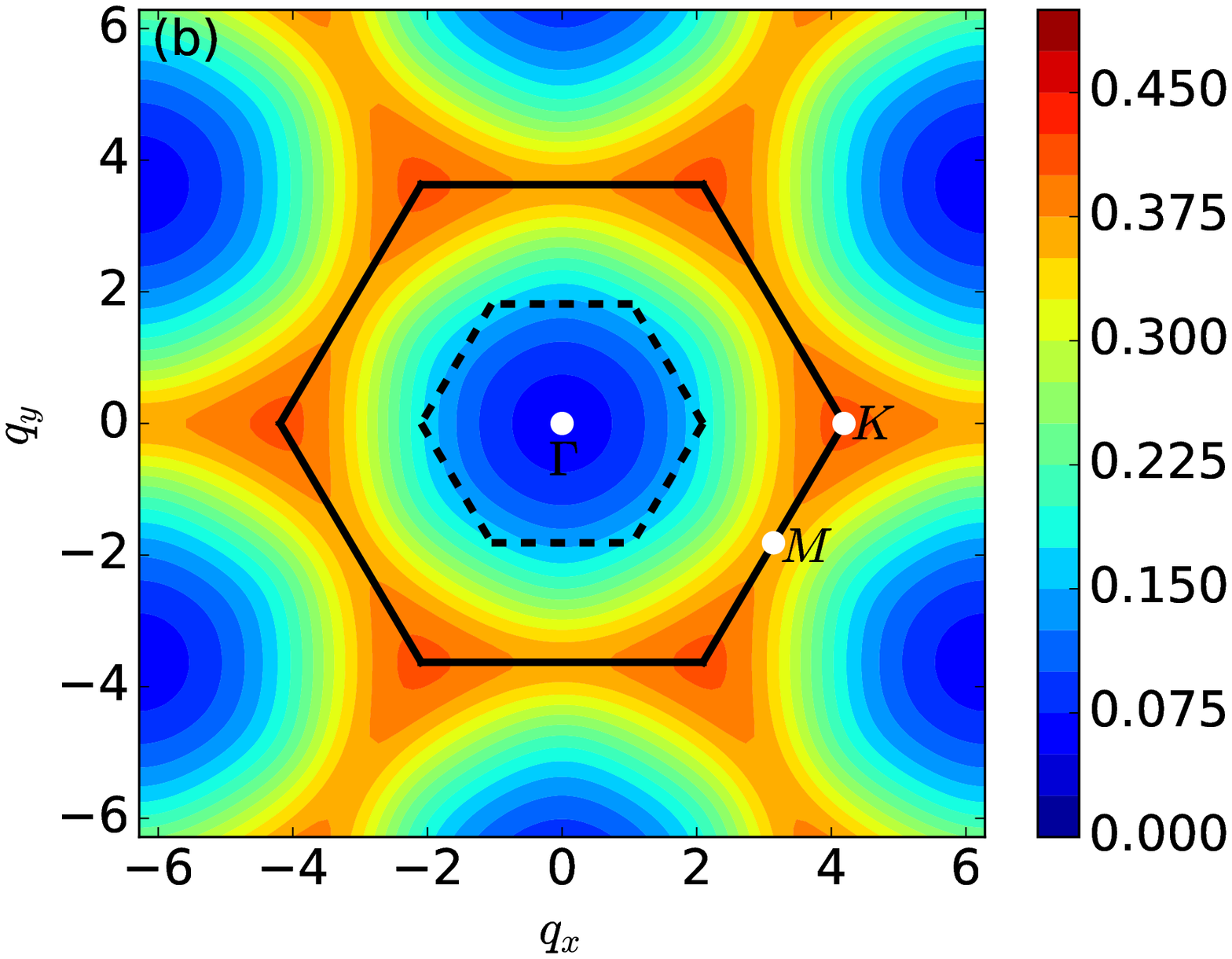}\\
\includegraphics[width=0.49\linewidth]{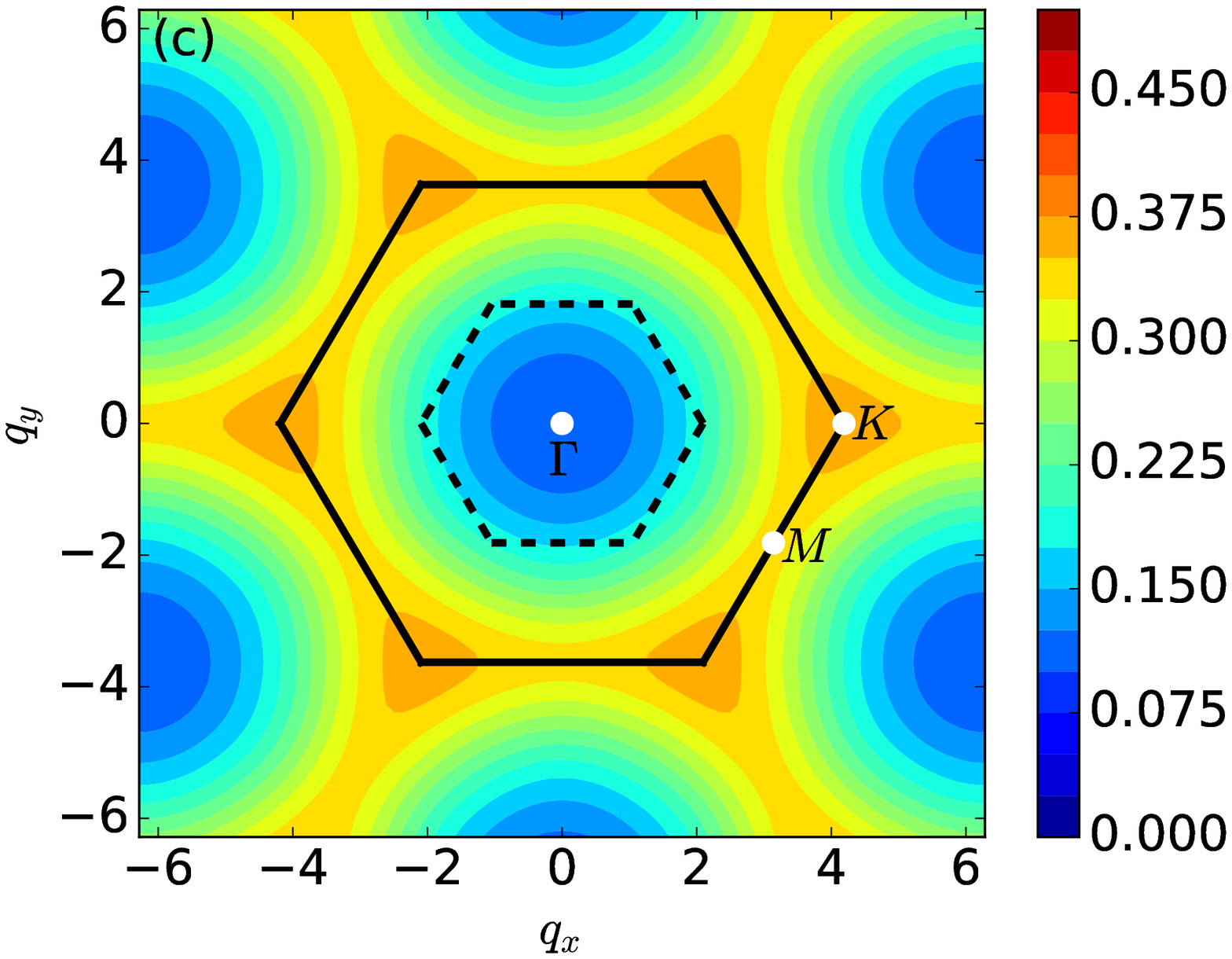}
\includegraphics[width=0.49\linewidth]{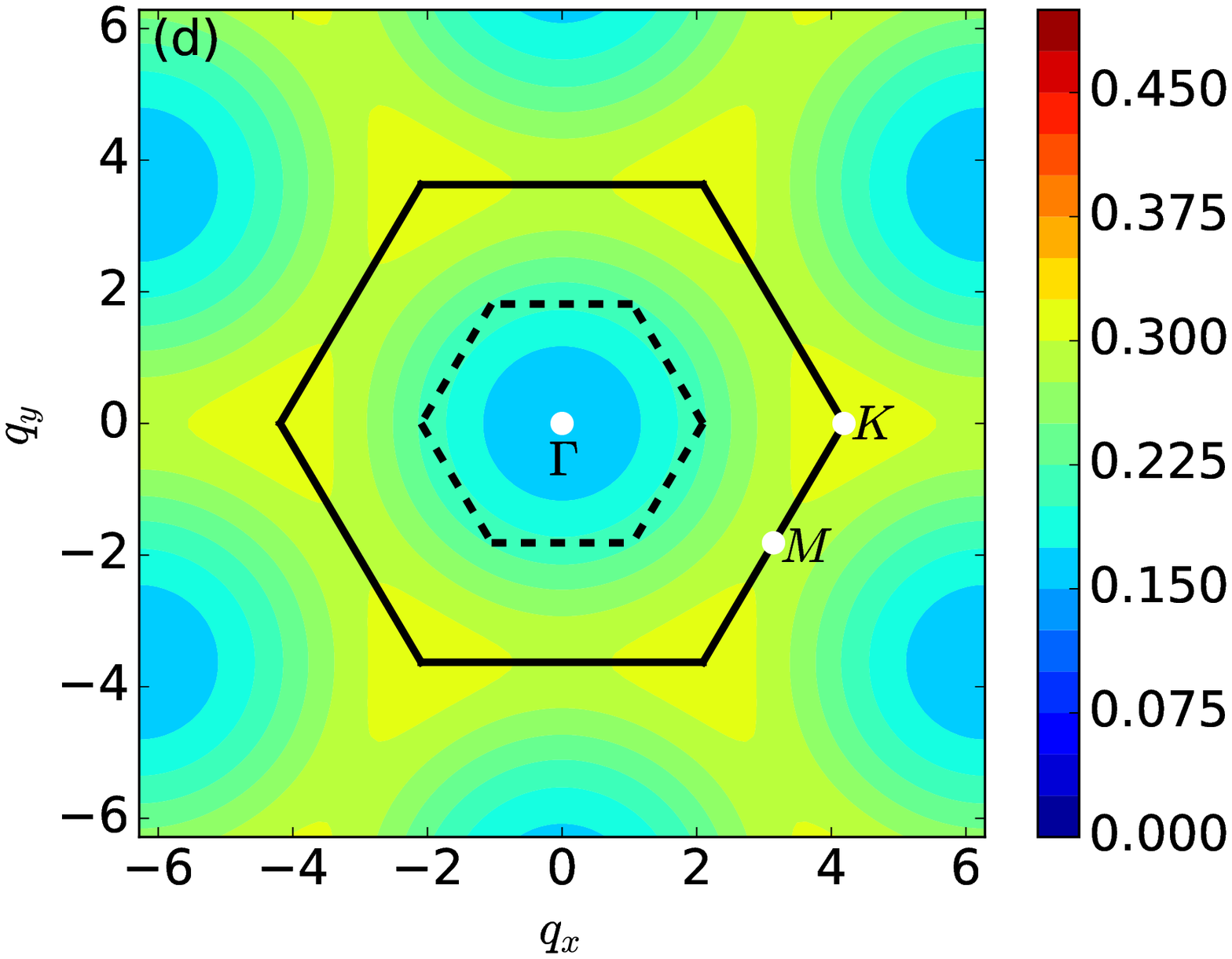}
\caption{\label{Sofq} The q-dependence of the equal time spin-spin correlation functions for the Kagome AFM Heisenberg model. The four plots depict the temperatures 0.25J, 0.5J, J, and 2J for plots (a), (b), (c), and (d) respectively.}
\end{center}
\end{figure}

Since these correlations die off so rapidly, it means terminating the Fourier transform over space at finite distances, as we have from only considering up to 7 triangles, is a very good approximation. The Fourier transform of the zeroth-moment yields the full q-dependence for the equal-time correlations as shown in Figure \ref{Sofq}. 
We see the characteristic development of dark nearly circular patches at the centers of the extended Brillouin Zone (BZ) at relatively high temperatures of
order $J$, and the intensity starts to concentrate on the boundaries of the extended BZ. 
Focusing on the bright zone boundary regions,
for all temperatures in this study, the maximum spectral weight is found at the K point, with a decrease in magnitude as we move towards the $M$ point, and a rapid drop-off as we move from the K point to the origin $\Gamma$.  

\begin{figure}
\begin{center}
\includegraphics[width=\linewidth]{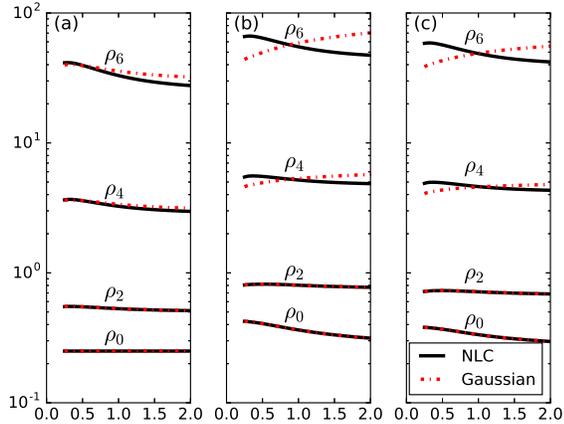}
\caption{\label{moments_compare} Comparison of the frequency moments of $S$ obtained directly from NLC, with those obtained in the gaussian approximation. Subplot (a) shows moments for $S(\vec{r}=0,\omega)$ ,  (b) and (c) show moments of $S(\vec{q},\omega)$ for the $K$ and $M$ wavevectors, as defined in Figure \ref{Sofq}, respectively.}
\end{center}
\end{figure}

We calculate the dynamic structure factors using the Gaussian approximation. To assess the limitations of this approximation,
we show in Figure \ref{moments_compare} a comparison between higher moments obtained from NLC with those obtained by the gaussian approximation. 
We show this comparison for $S(\vec{r}=0,\omega)$, as well as $S(\vec{q},\omega)$ for the $K$ and $M$ points, as defined in Figure \ref{Sofq}. We find that for the on-site calculation the gaussian approximation reproduces the higher moments very well and gives
a good approximation over the temperature range. However, for the $K$ and $M$ points, we find that the deviations from
gaussianity changes sign at a temperature below $J$. At high temperature the skew is towards lower frequencies, where as it
develops a high-frequency asymmetry at lower temperatures.

\begin{figure}
\begin{center}
\includegraphics[width=0.49\linewidth]{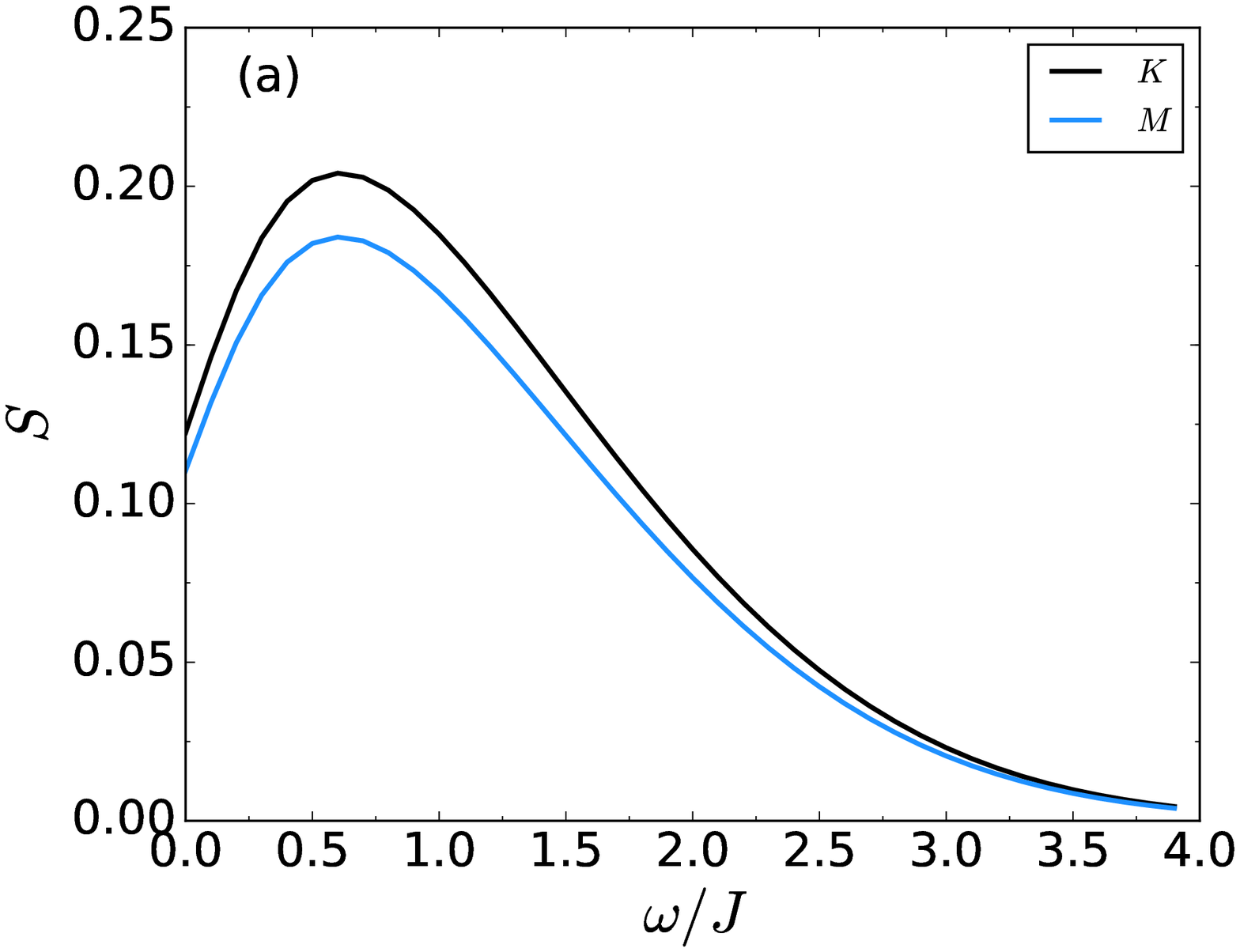}
\includegraphics[width=0.49\linewidth]{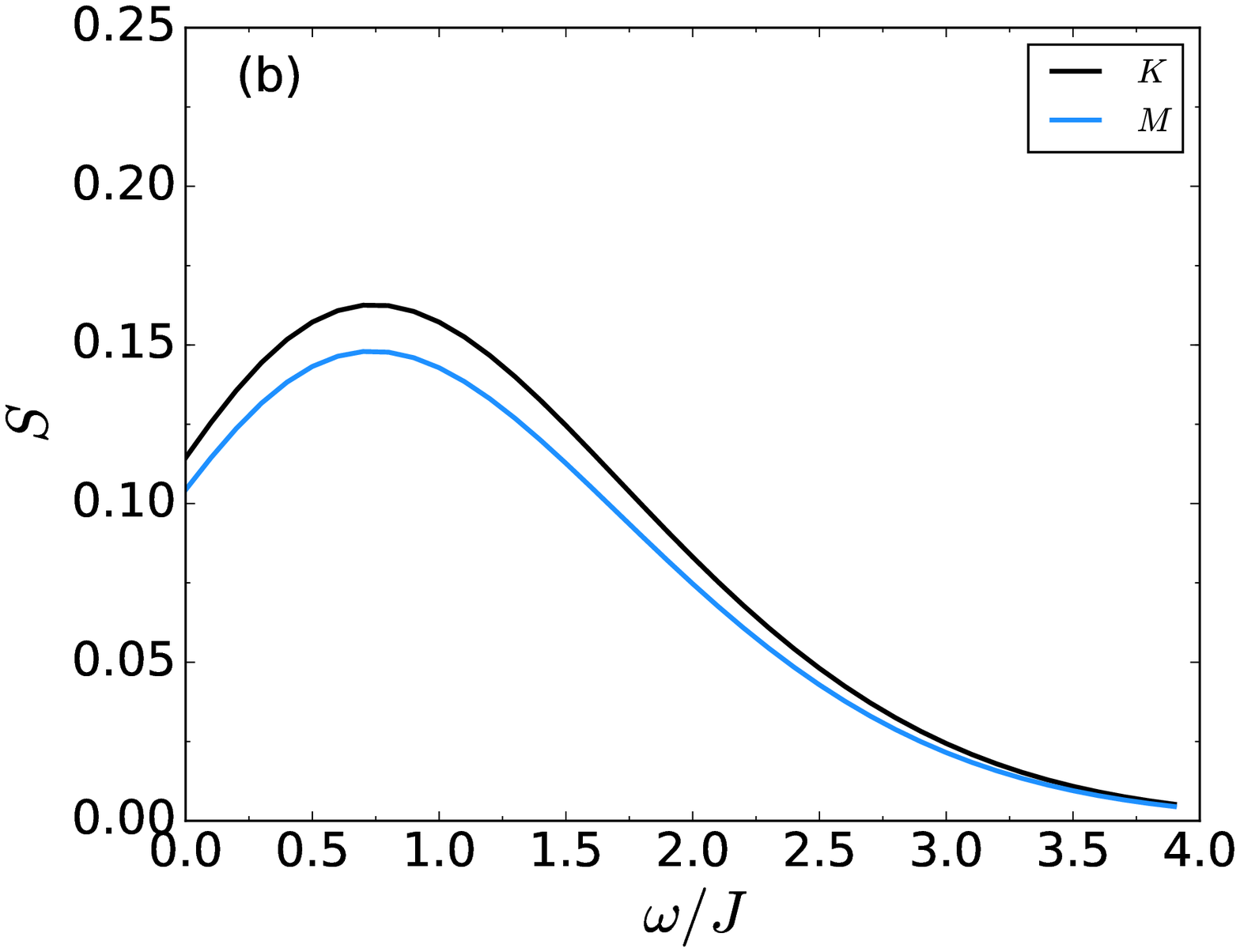}\\
\includegraphics[width=0.49\linewidth]{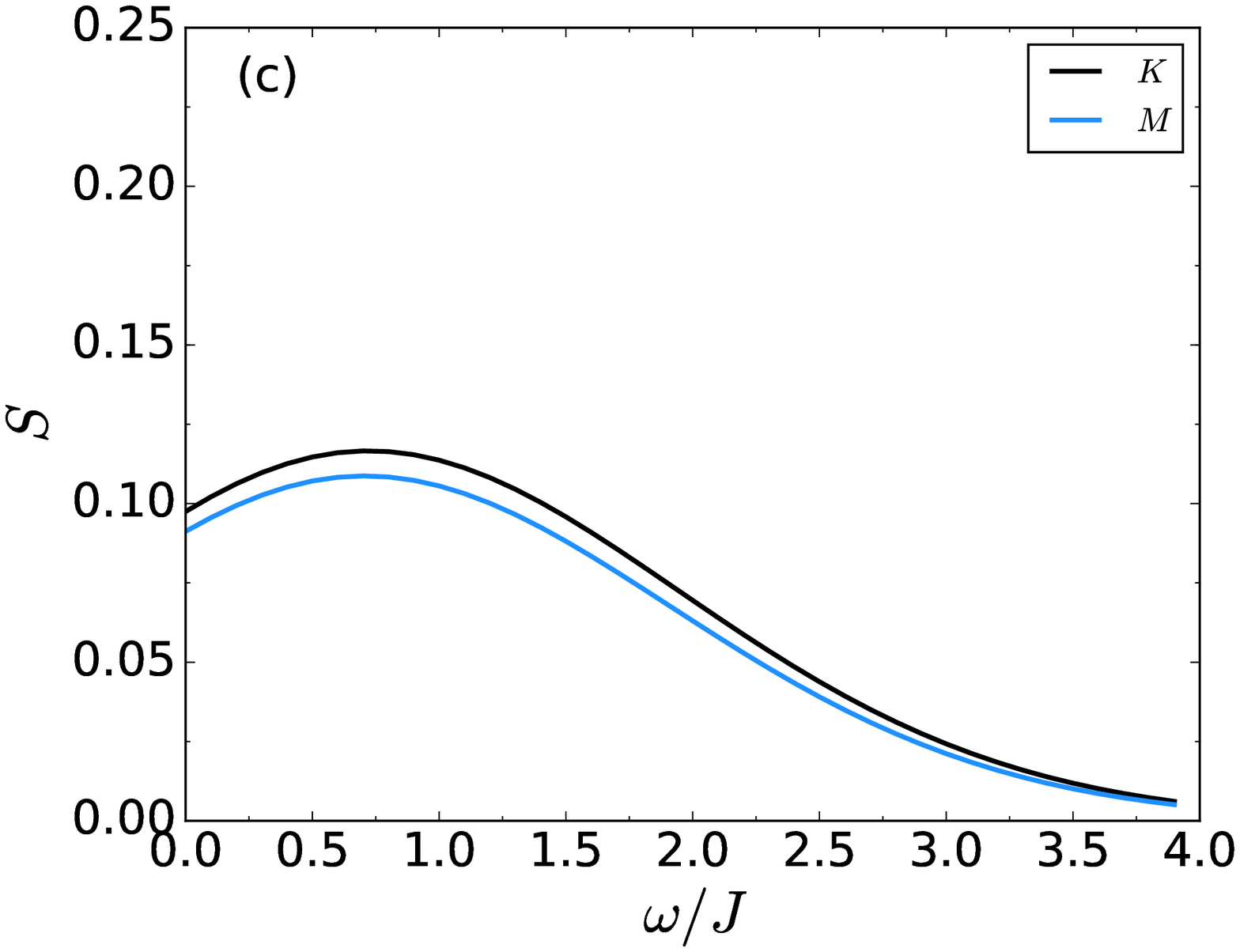}
\includegraphics[width=0.49\linewidth]{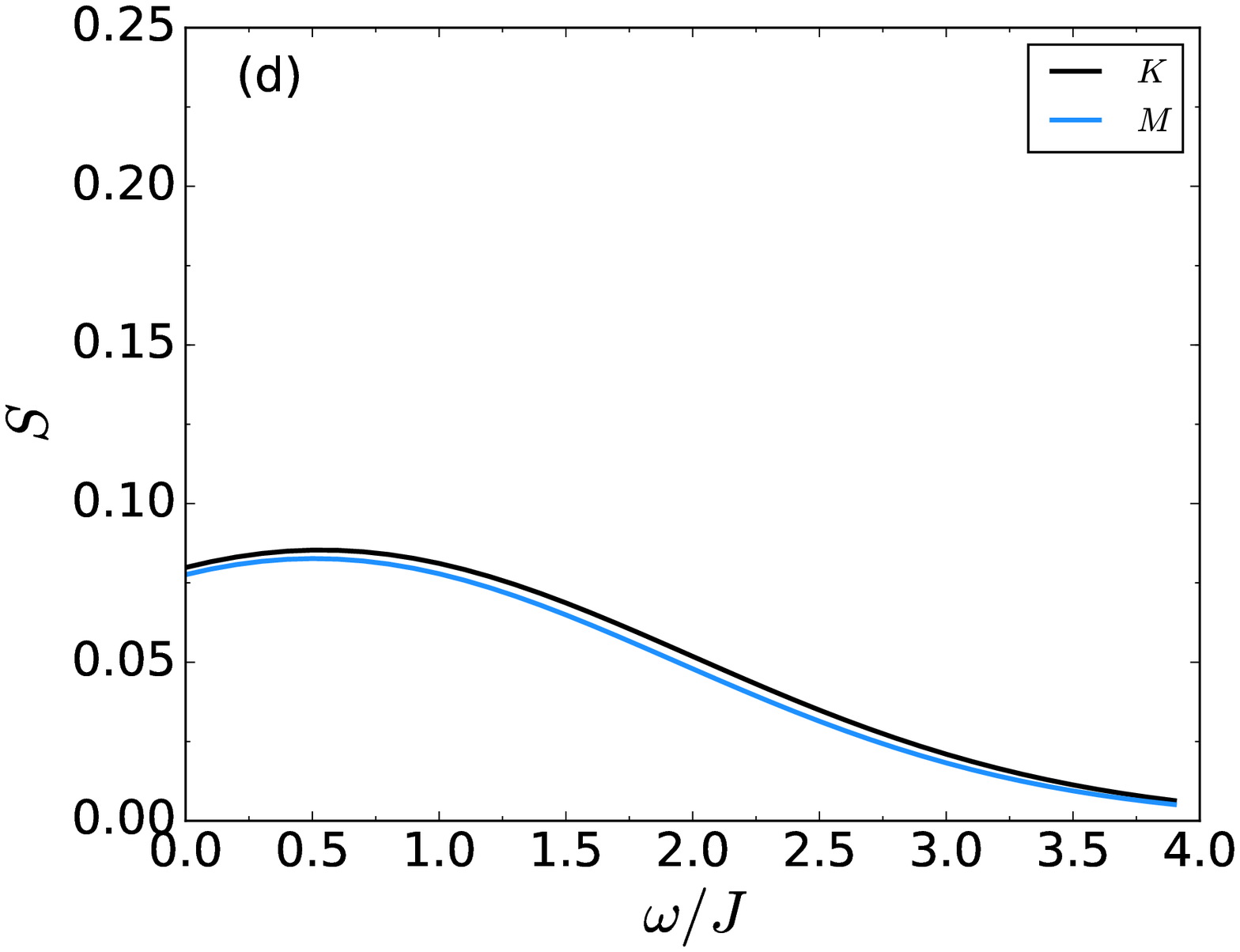}
\caption{\label{SQ-SofW} The frequency dependence of $S(q,\omega)$ within the Gaussian approximation for the Kagome AFM Heisenberg model. We show the K and M points as defined in Figure \ref{Sofq}. The four plots depict the temperatures 0.25J, 0.5J, J, and 2J for plots (a), (b), (c), and (d) respectively.}
\end{center}
\end{figure}

\begin{figure}
\begin{center}
\includegraphics[width=0.49\linewidth]{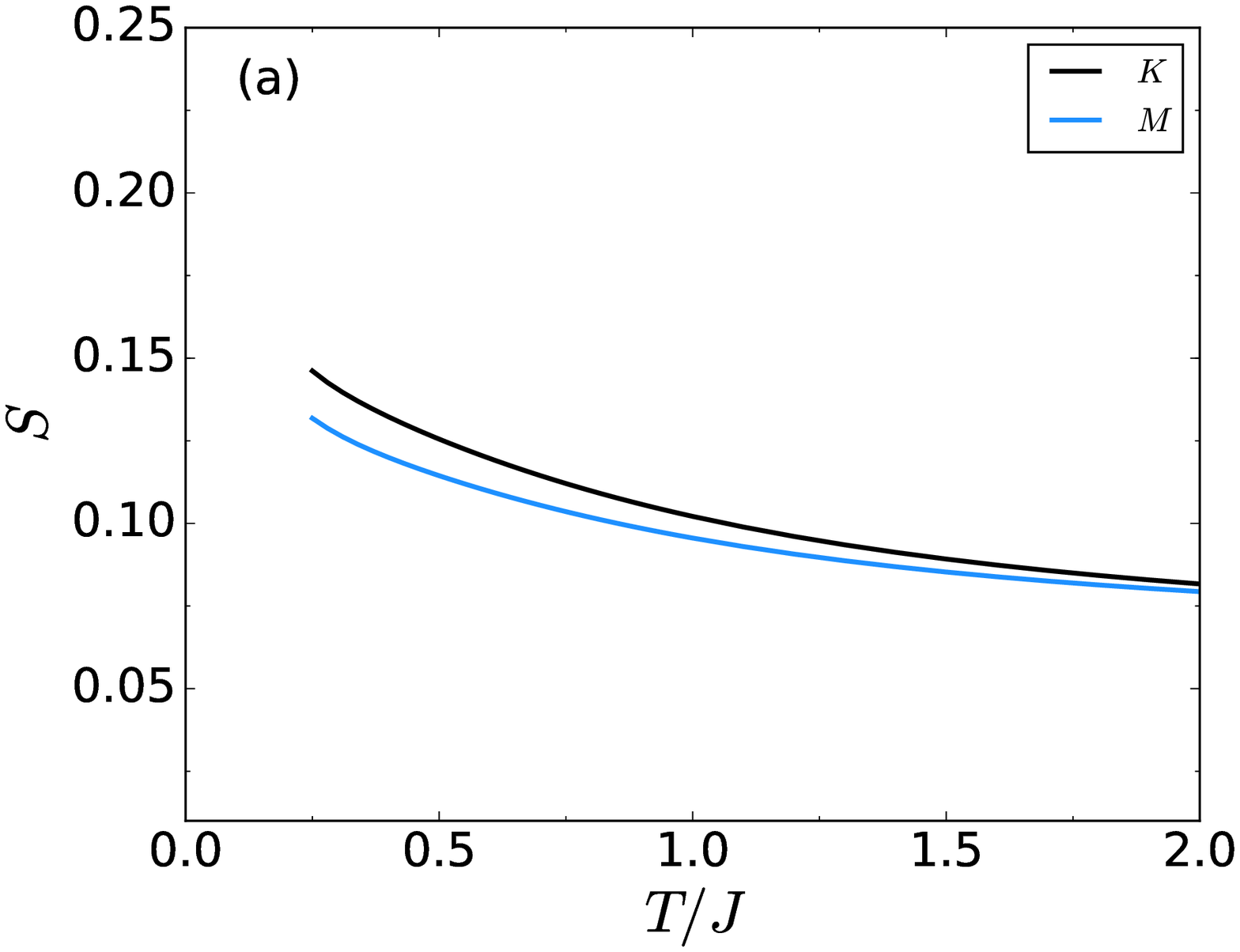}
\includegraphics[width=0.49\linewidth]{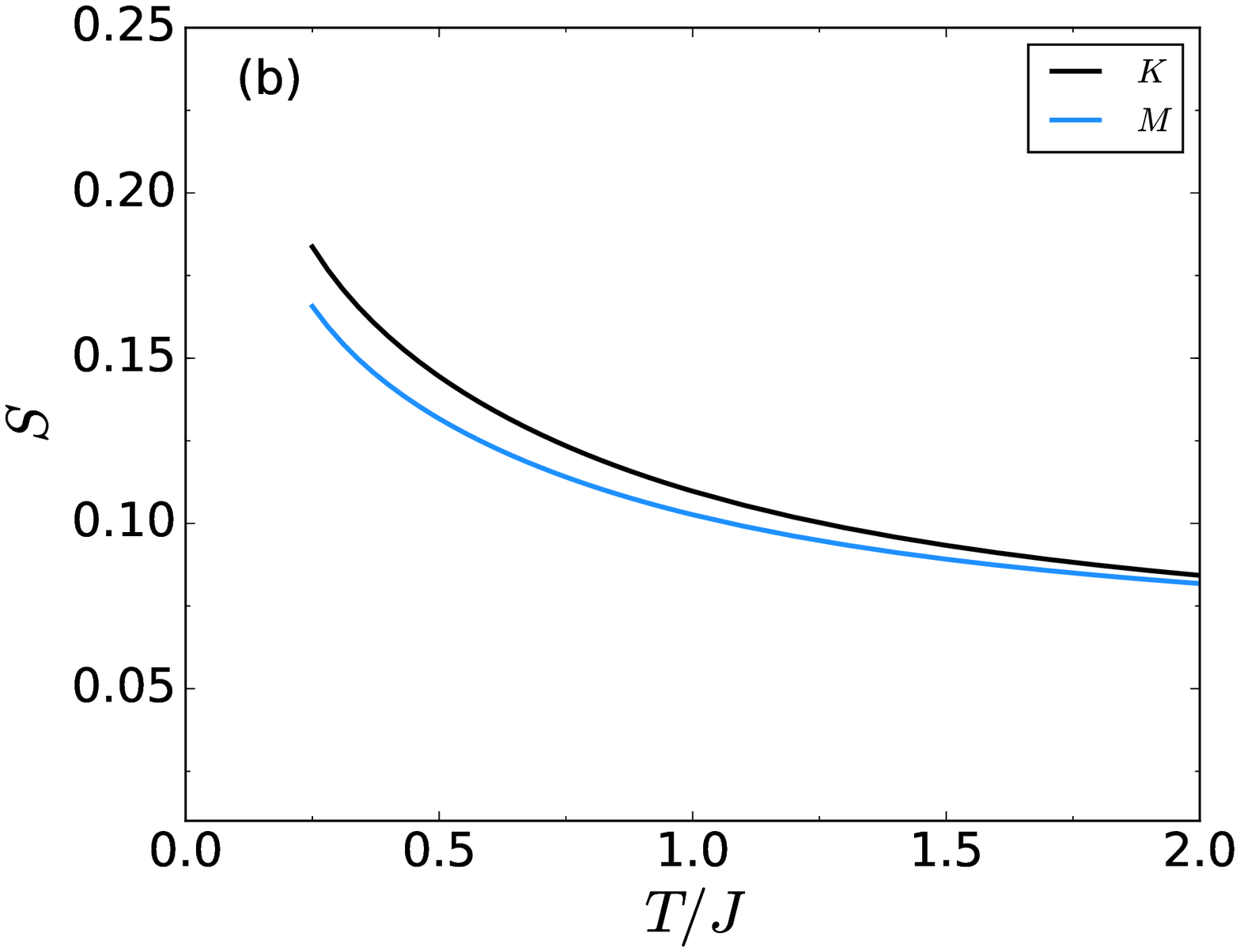}\\
\includegraphics[width=0.49\linewidth]{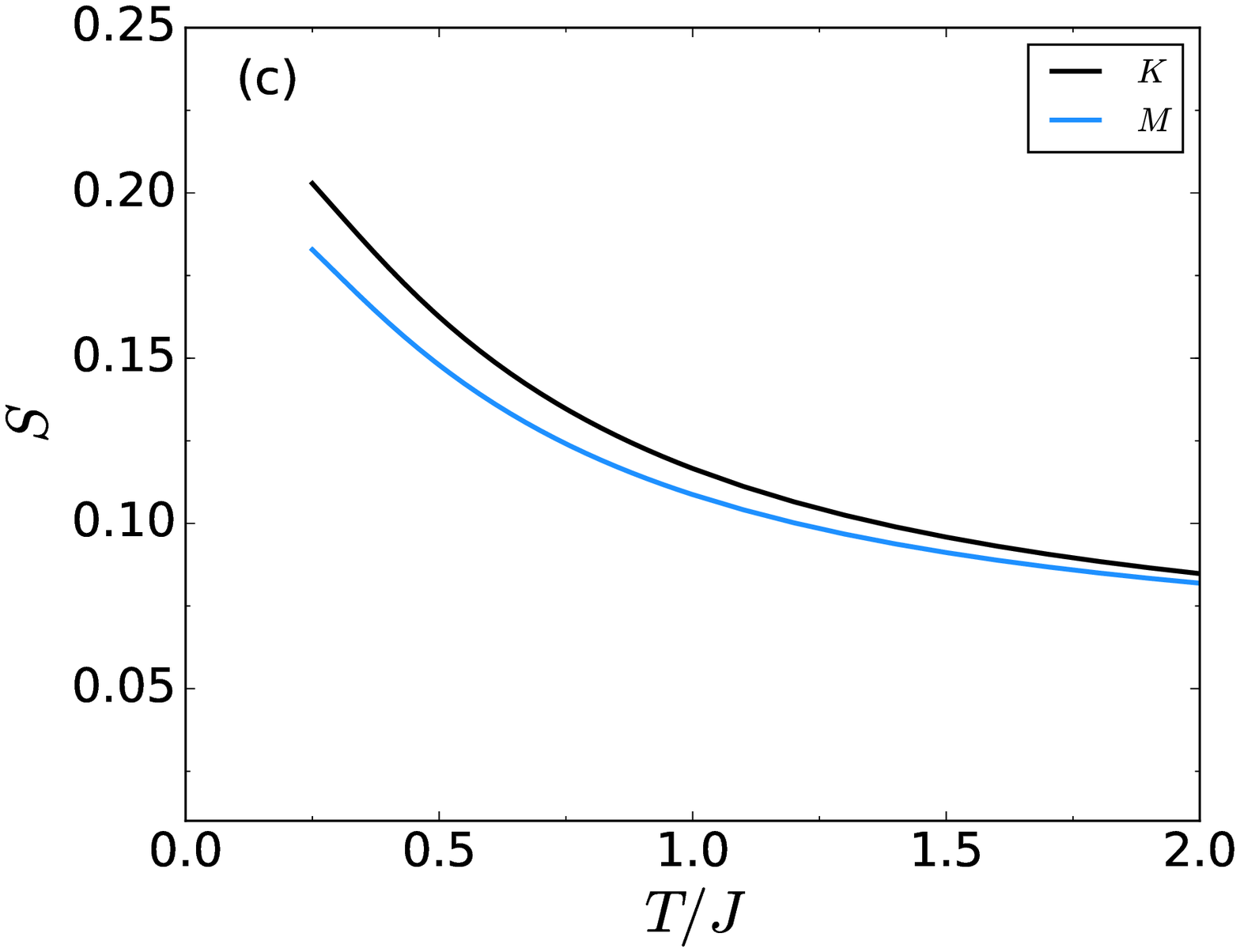}
\includegraphics[width=0.49\linewidth]{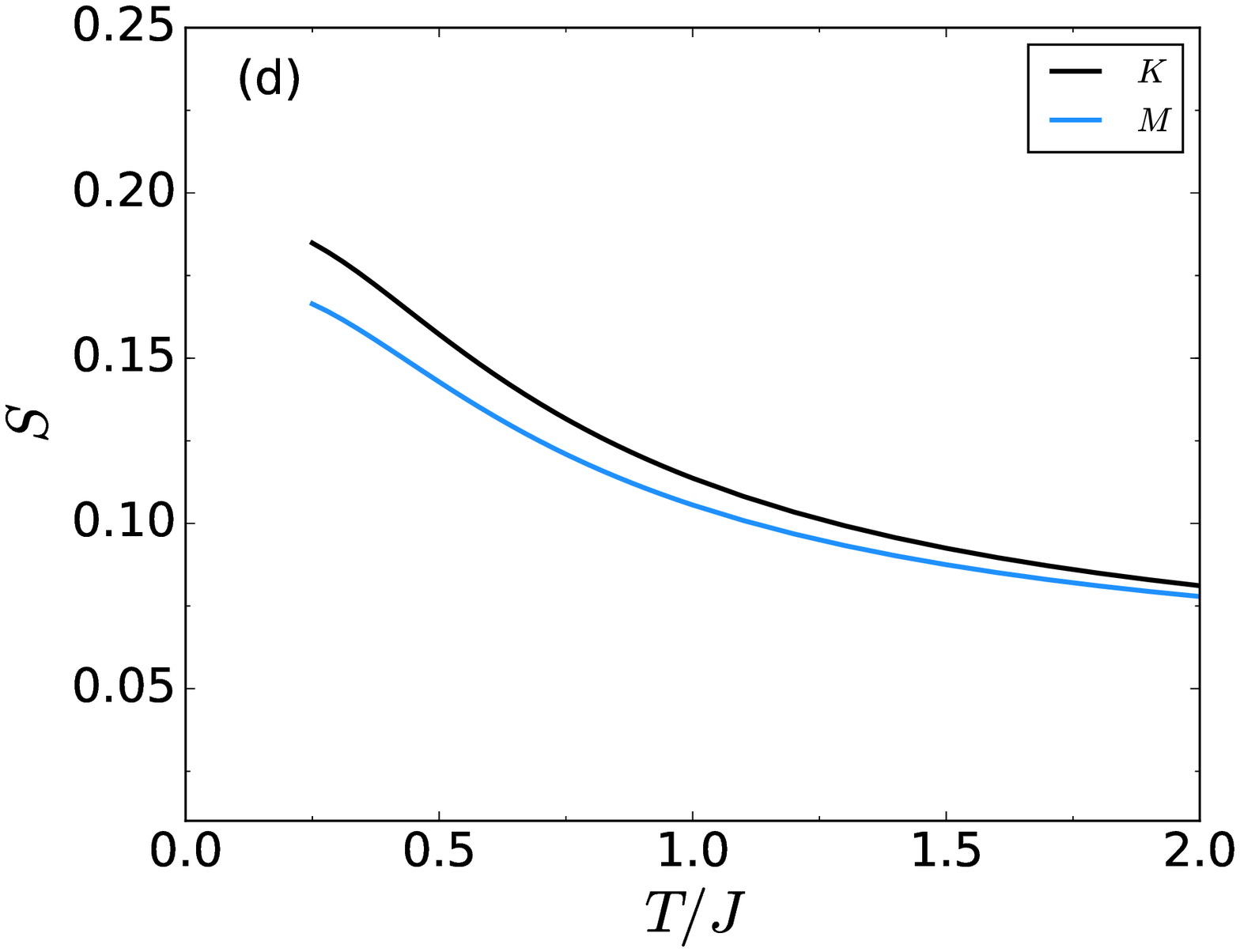}
\caption{\label{SQ-SofT} The temperature dependence of $S(q,\omega)$ within the Gaussian approximation for the Kagome AFM Heisenberg model. We show the K and M points as defined in Figure \ref{Sofq}. The four plots depict the frequencies 0.25J, 0.5J, J, and 2J for plots (a), (b), (c), and (d) respectively.}
\end{center}
\end{figure}

For the dynamic structure factor in the gaussian approximation, the intensity accumulates most prominently at the K point, and the line towards the $M$ point from K holds the most spectral weight. 
The frequency dependence for the K and M points for several temperature values 
is shown in Figure \ref{SQ-SofW}.
The intensity peaks around $\omega=0.7 J$.  On general grounds, one
expects the spectral weights to decrease rapidly above $\omega=2J$ \cite{nmr-theory}. That rapid decrease is ensured by the
Gaussian approximation.
For the same q values, we show the temperature dependence of $S(q,\omega)$ for several values of $\omega$ in Figure \ref{SQ-SofT}.
The intensity grows monotonically with decreasing temperature for the temperatures shown, and the overall behavior is
very similar to the equal-time correlation functions. 


\begin{figure}
\begin{center}
\includegraphics[width=\linewidth]{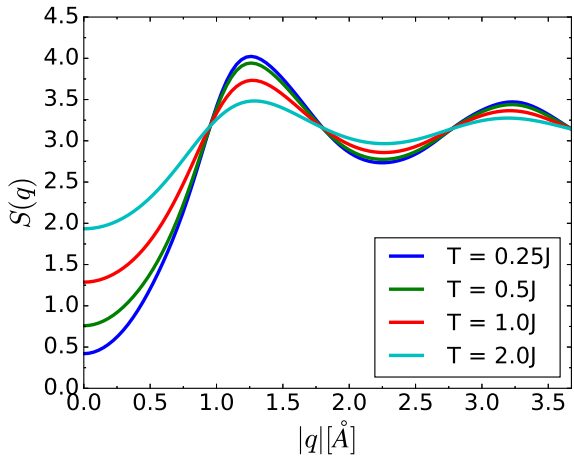}
\caption{\label{SQ-Polar} The q-dependence of the powder-average equal time spin-spin correlation functions for the Kagome AFM Heisenberg model, where we integrated over all values with the same $|q|$.}
\end{center}
\end{figure}

\begin{figure}
\begin{center}
\includegraphics[width=0.49\linewidth]{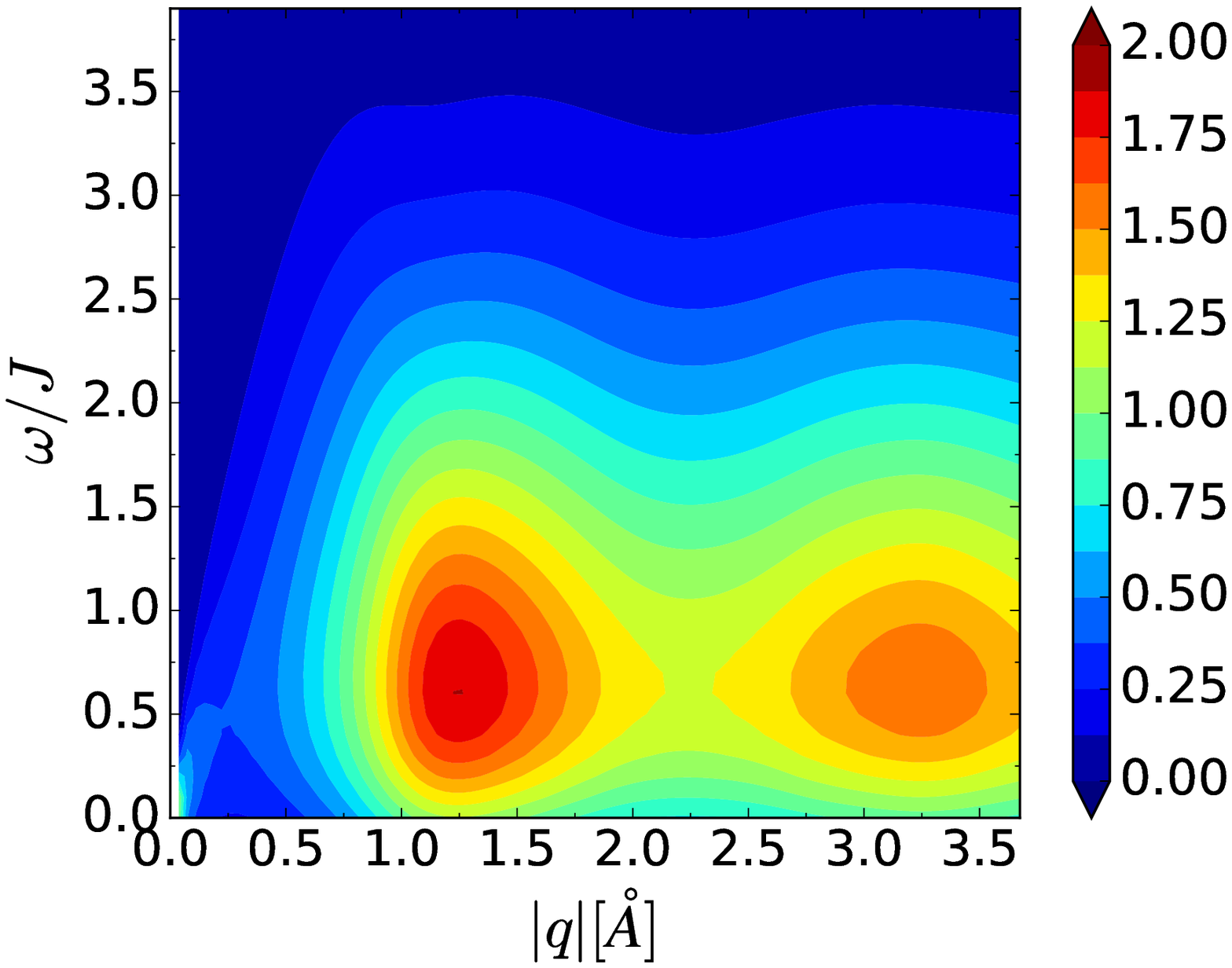}
\includegraphics[width=0.49\linewidth]{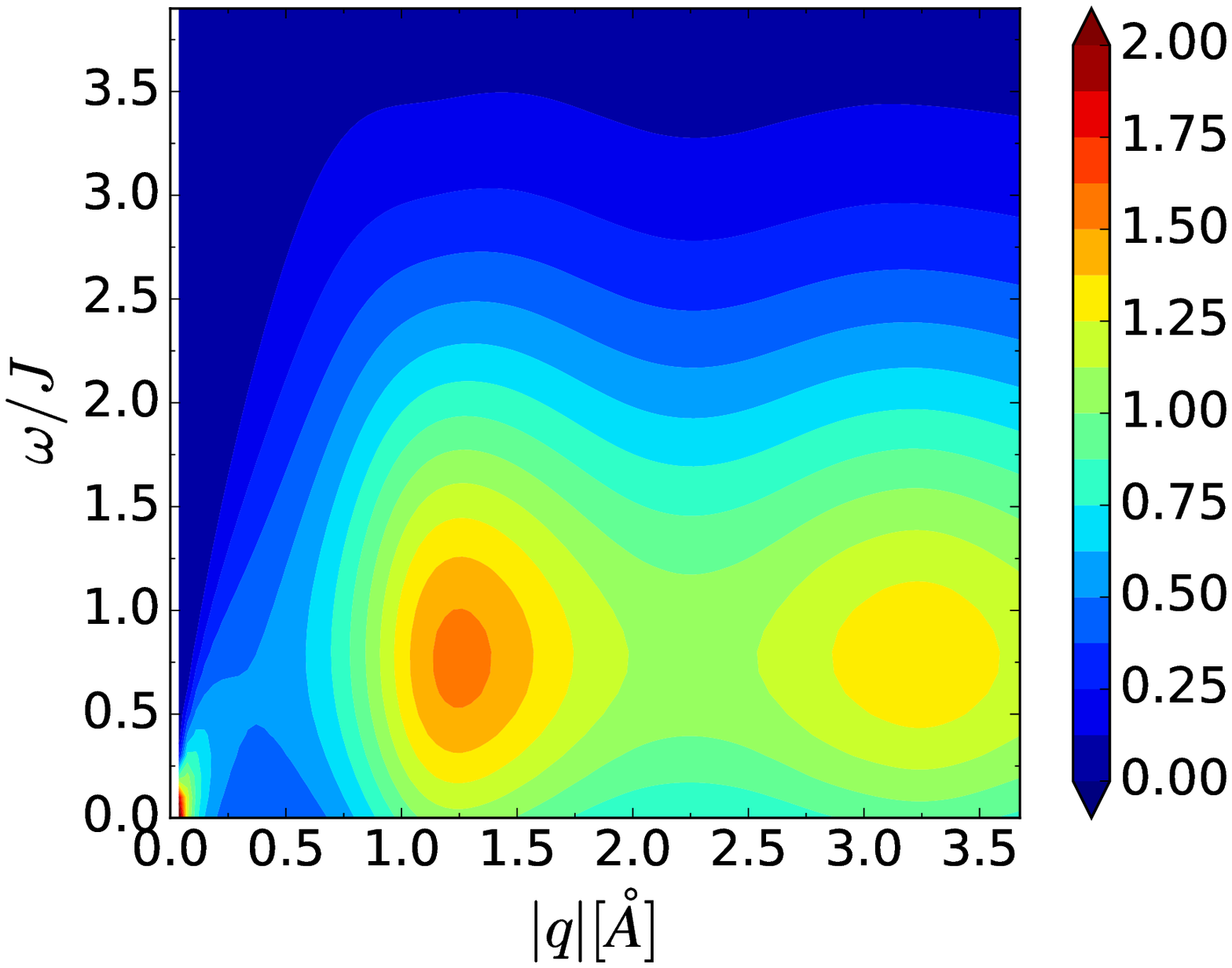}\\
\includegraphics[width=0.49\linewidth]{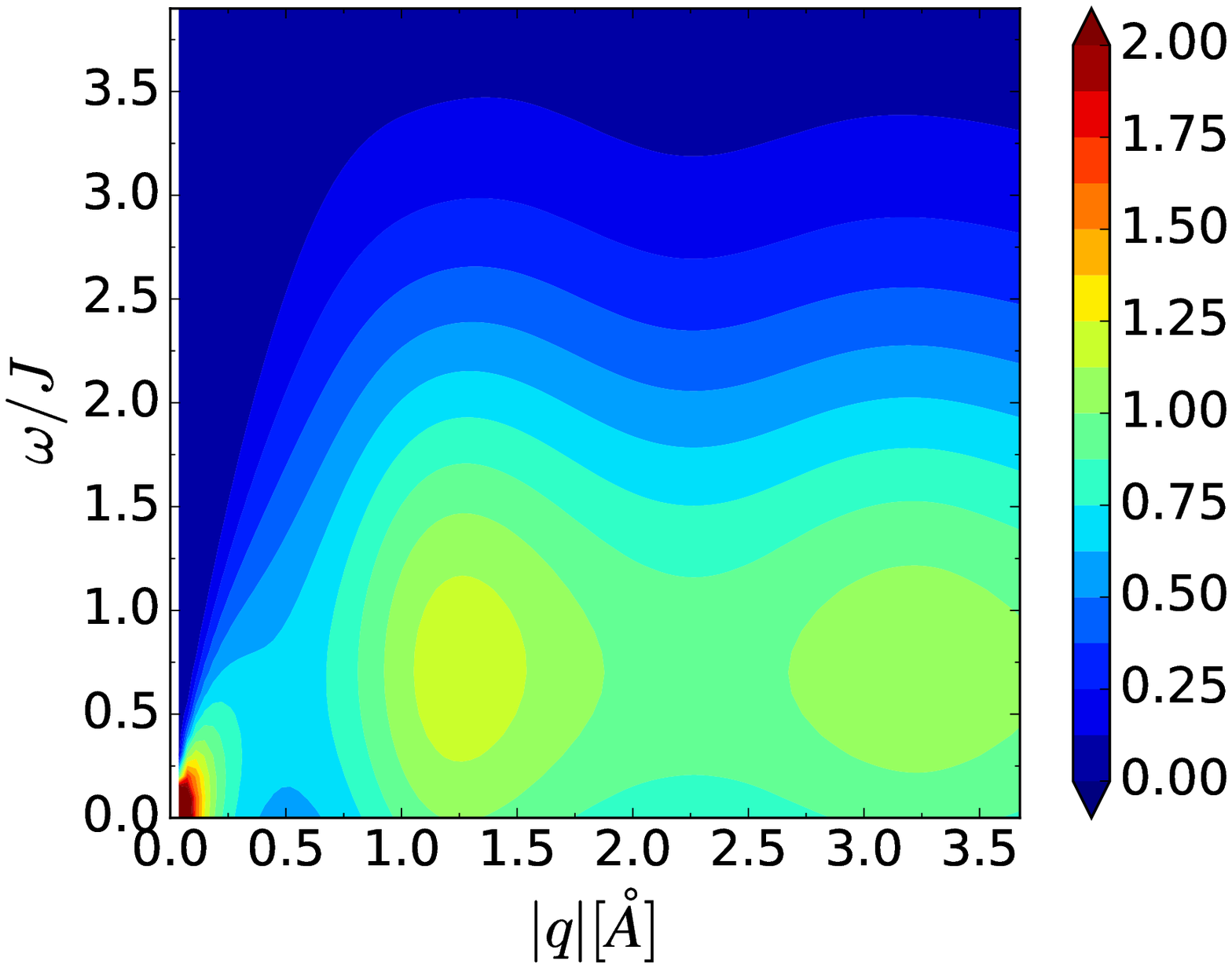}
\includegraphics[width=0.49\linewidth]{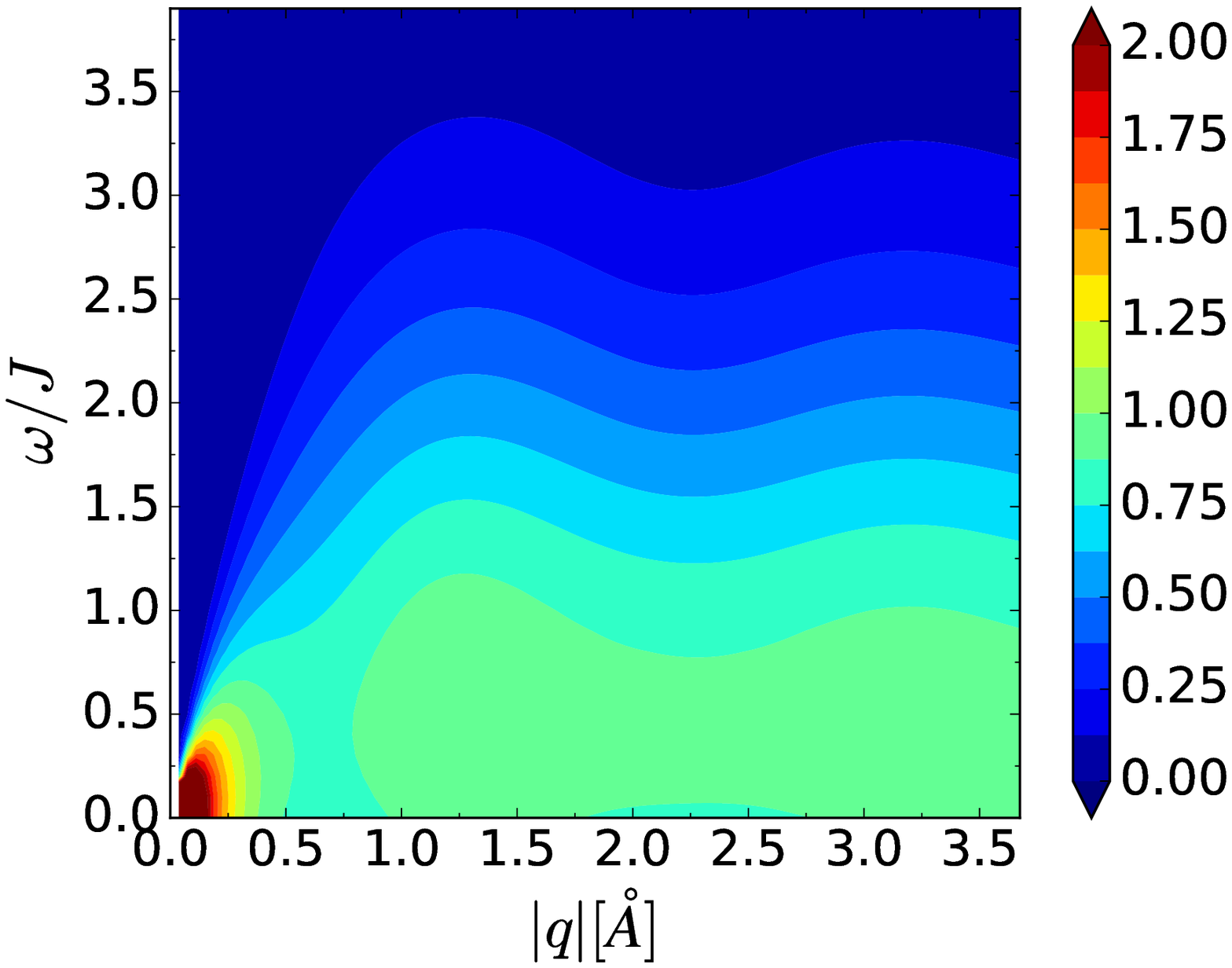}
\caption{\label{SQW-Polar} Powder average structure factors as a function of $q$ and $\omega$ within Gaussian approximation for the Kagome AFM Heisenberg model. The four plots depict the temperatures 0.25J, 0.5J, J, and 2J for plots (a), (b), (c), and (d) respectively.}
\end{center}
\end{figure}

We also compare directly with powder experiments by integrating over all points at equal $|\vec{q}|$.  
Here it is important to perform a 3D powder average as appropraite for the experiments of de Vries et al. \cite{Vries}
The angular averages reduce to
\begin{equation}
[ \exp{ i \vec q \cdot \vec r}]_{av} = \frac{\sin{qr}}{qr}.
\end{equation}
Thus, they are easy to obtain from the real-space correlations.
Assuming the lattice spacing between neighboring copper ions in Herbertsmithite is $a=3.4\AA$ as stated by de Vries {\it et al} \cite{Vries}, we show $S(q)$ vs $|q|$ in Figure \ref{SQ-Polar}. We find peaks at $|q|\approx 1.3 \AA^{-1}$ and $|q|\approx 3.2 \AA^{-1}$, and a trough near $|q|=2.2 \AA^{-1}$, in very good agreement with the experiments \cite{Vries} which found a peak at $|q| \approx 1.3\AA^{-1}$. 
We also show plots of powder-average $S(q,\omega)$ for various temperatures in Figure \ref{SQW-Polar}.  
We show a range of values of $|q|$, and find peaks developing at $|q|\approx1.1,3.0$ and $\omega\approx0.6$ and this intensity diminishes in all direction in the $\omega-|q|$ plane from there.

\section{Discussions and Conclusions}
In this paper, we have calculated the real-space spin correlations and wavevector and frequency dependent structure factors
of the Kagome Lattice Heisenberg Model (KLHM) at finite temperatures using the Numerical Linked Cluster (NLC) method.
These calculations should be very accurate for the static structure factors, in the thermodynamic limit, for $T>J/4$. The frequency dependence is
obtained using the Gaussian approximation maintaining the fluctuation-dissipation relations. 

The development of short-range antiferromagnetic order sets in at temperatures of order $J$ and leads to
drak patches at the extended Brillouin Zone centers and enhanced
spectral weights along the boundaries of the extended Brillouin Zone. Our results for powder diffraction are in
very good agreement with the Neutron spectra on the Herbertsmithite materials
obtained earlier by de Vries et al \cite{Vries}. Earlier NMR relaxation rates, which
are a sum over wavevectors, calculated
using the NLC, were also found to be in good agreement with experiments \cite{nmr-theory}.

Recently full wavevector resolved neutron spectra were measured on single crystals of Herbertsmithites
at low temperatures ($T=J/100$) \cite{neutron12}. We are not aware of any such measurements at higher temperatures.
Comparison of our calculated spectra at much higher temperatures than the experiments show agreement
with the broad features where spectral weight begins to get concentrated at the boundaries of the
extended Brillouin Zone. However, our results differ from the experiments in one rather striking regard.
We find the intensity maximum at the $K$ point on the corners of the extended Brillouin Zone. In contrast,
the low temperature experiments show a peak at the $M$ points at the middle of the boundary of the extended
Brillouin Zone. 
We suggest that these measurements should be done as a function of temperature. In the
absence of a crossover of a maximum from the $M$-point to the $K$-point as a function of temperature \cite{kawamura,hao}, the Herbertsmithite materials
must differ from the KLHM in some important regard. However, if such a crossover is found, its temperature
will provide an important crossover energy scale for the material. 

From a theoretical point of view, a peak in the structure factor at the $K$ point is consistent with order
in the $\sqrt{3}\times \sqrt{3}$ pattern, which is favored in classical, large-S and many computational 
approaches\cite{hao,harris,rutenberg,chernyshev,richter,oitmaa} , 
where as a peak at the $M$ point is consistent with order at $q=0$.
As the computational studies show, these two classical patterns are very close in energy but the spin-half Heisenberg
model does not have long-range order in the ground state at all. Nevertheless, there may be competition for short-range order
reflected in the dominance of different q points \cite{sachdev}. 
The higher temperature behavior favors the semiclassical and perturbative
approaches. But, the ground state studies of the largest clusters studied show otherwise \cite{kawamura,lauchli}. 
A recent theoretical work discusses a crossover from a conventional spin-liquid to an algebraic spin-liquid at
a temperature above $J/2$ \cite{xichen}.
It would be interesting to have experimental input to the competition and crossover between these ordering tendencies, 
which would require the wave-vector resolved measurements to be done at higher temperatures.

\section{Acknowledgements}
We would like to thank Jeff Rau for a careful reading of the manuscript.
This work is supported in part by the National Science Foundation grant number DMR-1306048.

\end{document}